\documentclass[journal=jctc,manuscript=article]{achemso}
\usepackage{graphicx}
\usepackage{rotating}
\usepackage{amsfonts,amsmath,amssymb,latexsym,bbm,dsfont}
\usepackage{chemformula}
\usepackage{color}
\usepackage{subcaption} 
\usepackage{booktabs}
\usepackage{geometry}
\usepackage{multirow}
\usepackage{float}
\usepackage{url}

\title{Transcorrelated Theory with Pseudopotentials}

\author{Kristoffer Simula}
\affiliation{Max Planck Institute for Solid State Research, Heisenbergstr. 1, 70569 Stuttgart, Germany}
\email{k.simula@fkf.mpg.de}

\author{Evelin Martine Corvid Christlmaier}
\affiliation{Max Planck Institute for Solid State Research, Heisenbergstr. 1, 70569 Stuttgart, Germany}

\author{Maria-Andreea Filip}
\affiliation{Max Planck Institute for Solid State Research, Heisenbergstr. 1, 70569 Stuttgart, Germany}

\author{J. Philip Haupt}
\affiliation{Max Planck Institute for Solid State Research, Heisenbergstr. 1, 70569 Stuttgart, Germany}

\author{Daniel Kats}
\affiliation{Max Planck Institute for Solid State Research, Heisenbergstr. 1, 70569 Stuttgart, Germany}

\author{Pablo Lopez-Rios}
\affiliation{Max Planck Institute for Solid State Research, Heisenbergstr. 1, 70569 Stuttgart, Germany}

\author{Ali Alavi}
\affiliation{Max Planck Institute for Solid State Research, Heisenbergstr. 1, 70569 Stuttgart, Germany}
\alsoaffiliation{Yusuf Hamied Department of Chemistry, University of Cambridge, Lensfield Road, Cambridge CB2 1EW, United Kingdom}
\email{a.alavi@fkf.mpg.de}
\date{\today}

\begin{document}

%\maketitle

\begin{abstract}
    The transcorrelated (TC) method performs a similarity transformation on the
    electronic Schrödinger equation via Jastrow factorization of the wave function. This
    has demonstrated significant advancements in computational 
    electronic structure theory by  
    improving basis set convergence and compactifying the description of the wave function. In 
    this work, we introduce a new approach that incorporates pseudopotentials (PPs) into the 
    TC framework, significantly accelerating Jastrow factor optimization and reducing 
    computational costs. Our results for ionization potentials, atomization energies, and 
    dissociation curves of first-row atoms and molecules show that PPs provide chemically 
    accurate descriptions across a range of systems and give guidelines for future 
    theory and applications.
    The new pseudopotential-based TC method opens possibilities for applying TC to more complex 
    and larger systems, such as transition metals and solid-state systems.
\end{abstract}

\begin{tocentry}
  \centering
  % if PDF:
  \includegraphics[width=\linewidth]{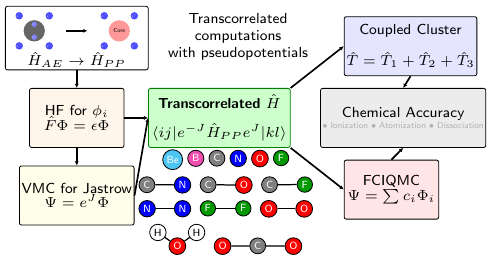}
  % or if TIFF:
  % \includegraphics[width=\linewidth]{toc_graphic.tif}
\end{tocentry}

\section{Introduction}

In methods based on electronic structure theory, solutions 
to the Schrödinger equation require 
simultaneous convergence with respect to basis sets and the treatment of 
electron correlation. 
Current \textit{ab initio} methods that treat the electron correlation explicitly
 can achieve this accurately only for very 
small systems, due to the high polynomial or even exponential scaling of the
computational cost with the 
number of electrons.

First quantized methods, such as Variational and Diffusion Monte Carlo (VMC and 
DMC), work in continuum space, circumventing the basis set 
issue. However, their accuracy is limited by the quality of the trial wave 
function. Systematic improvement of the wave function requires 
methods capable of treating electron correlation effectively.

Conversely, methods such as Coupled Cluster (CC) and 
configuration interaction (CI), formulated under the framework of second quantization,
can often treat electron correlation accurately, 
but convergence to the basis set limit remains infeasible in many cases. 
Additionally, the divergent nature of the Coulomb interaction introduces sharp 
features, or cusps \cite{kato1957}, in the wave function, which are difficult to 
describe using determinants composed of smooth Gaussian-type orbitals, which 
are often used in second-quantized post-Hartree-Fock methods.

These issues of second quantization have been alleviated by explicitly correlated methods, which 
introduce interelectronic distance dependency in the system description. This is done 
to reduce basis set errors and account for the cusps, and to describe short-range 
electronic interactions without requiring a large number of determinants.

R12/F12 methods \cite{kutzelnigg1985,ten2004} have been widely applied in quantum 
chemistry to address these problems, and have been used in perturbation theory, 
coupled cluster, and configuration interaction calculations \cite{kutzelnigg1991,
noga1994,kong2012}, generally with good results. 

The transcorrelated (TC) method is an explicitly correlated method 
that has seen rapid development in recent years \cite{cohen2019,ammar2023,ammar2023_2,lee2023,ten2023, ammar2024}. 
TC takes a Jastrow factor, a function of 
interelectronic distances optimized in a first-quantized VMC calculation, and uses it
to perform a similarity transformation on the second-quantized Hamiltonian. 
This transformation preserves the Hamiltonian’s eigenvalues while addressing the cusps 
and significantly improving basis set convergence \cite{haupt2023}, as well as 
compactifying the wave function \cite{dobrautz2019,ammar2024}. Although the transformation 
introduces challenging three-body terms, a recent approximation, xTC 
\cite{christlmaier2023}, removes the need to explicitly treat these terms, reducing the 
scaling in evaluation of the transcorrelated Hamiltonian by two orders of magnitude.

TC has shown very promising results in homogeneous electron gas (HEG) systems \cite{luo2018,liao2021},
and the Hubbard model \cite{dobrautz2019}. It has also been applied to atoms and
molecules with high accuracy using FCIQMC and CC methods \cite{cohen2019,guther2021,schraivogel2021,schraivogel2023}.
TC has also been used to produce accurate results in simulations with quantum
hardware, where the TC Hamiltonian enables the use of shallower circuit depths \cite{sokolov2023}.

One of the bottlenecks of TC and xTC has been the optimization of the Jastrow factor, 
as the variance of the VMC wave function increases rapidly with system size, driving 
up computational costs for sufficient accuracy. To extend TC and xTC to larger systems 
or solids, further developments are needed to reduce the costs of Jastrow optimization 
and the following post-Hartree-Fock calculations. The pseudopotential (PP) 
approximation offers a promising solution, as it reduces the number of electrons, 
eliminates electron-nucleus cusps, and significantly lowers the VMC variance.

In this work, we investigate the use of PPs in xTC methods. Replacement of the 
nucleus and the surrounding core electrons with PPs introduces terms in the 
Hamiltonian that do not commute with the Jastrow factor, complicating the similarity 
transformation. In Section 2, we present the theory of transcorrelation with 
PPs. In Section 3, we discuss the computational details of our calculations. 
Section 4 presents results on ionization potentials for first-row elements (Be–F), 
atomization energies for molecules (CN, CO, CF, N$_2$, O$_2$, F$_2$, H$_2$O, CO$_2$), 
and dissociation curves for N$_2$ and F$_2$. We show that PPs accelerate 
the optimization of the Jastrow factor and provide chemically accurate descriptions 
across a variety of systems and chemical environments.

\section{Theory}

\subsection{Pseudopotential approximation}

Within the PP approximation the Coulombic interactions between 
the valence electrons and the atomic nuclei and core electrons are replaced with
effective potentials. The PPs are constructed to reproduce the 
valence electron wave functions outside of a core region. This means that 
the electron-nucleus term of the Hamiltonian $\hat{H}_\text{en}$ is replaced by a 
sum over effective potentials $\hat{V}_\text{eff}$ for each atom:
\begin{equation}
    \label{eq:effective potential}
    \hat{H}_\text{en} = \sum_I^M\sum_{i=1}^{N}\frac{Z_I}{|\mathbf{r}_i-\mathbf{R}_I|} \rightarrow
    \hat{H}^\text{PP}_\text{en} = \sum_I^M\sum_{j=1}^{N_v}  \hat{V}^I_\text{eff}(|\mathbf{r}_j-\mathbf{R}_I|) .
\end{equation}
Above, $Z$ is the nuclear charge, $N$ ($M$) is the number of electrons (atoms), 
and $\mathbf{r}_i$ ($\mathbf{R}_I$) is the position of the $i$-th electron 
($I$-th atom) with respect to the origin. With PPs,
the summation runs over $N_v$ valence electrons.

In the theory that follows, we focus 
on the effective potential of a single atom ($M=1$) for to simplify presentation, but the extension 
to multiatom systems is straightforward.

Outside of the core region $V^I_\text{eff}$ should mimic the potential felt by the valence electrons due to the nucleus and the core electrons.
It should also
reproduce the exact electronic wave function outside of the core region specified 
by a cutoff radius $r_c$. $V_\text{eff}$ consists of a number of angular momentum channels, 
one local without spherical projections  and one or more nonlocal channels:
\begin{equation}
    \hat{V}_\text{eff}(r) =V_{l_{\text{max}}}(r) + \sum_{l=0}^{l_{\text{max}}-1}V_{l}(r) \sum_{m=-l}^{l}  |Y_{lm}\rangle \langle Y_{lm}| .
\end{equation} 
Above, $Y_{lm}$ are the spherical harmonics, and $V_{l}(r)$ are the pseudopotential radial functions.
$l_{\text{max}}$ is the maximum angular momentum quantum number included in the PP, which is also 
chosen as the local channel. $V_l$ are expressed as:
\begin{equation}
    V_{l}(r) = \left\{
    \begin{array}{ll}
        -\frac{Z_\text{eff}}{r}\left( 1-e^{-\alpha r^2}\right) + \alpha Z_\text{eff} re^{-\beta r^2} + \sum_{q=1}^{n}\gamma_{ql}e^{-\delta_{ql}r^2} & l=l_{\text{max}} \\
        \sum_{q=1}^{m}\gamma_{ql}e^{-\delta_{ql}r^2} & l < l_{\text{max}}
    \end{array}
    \right.
\end{equation}

\begin{figure}[H]
    \centering
    \includegraphics[scale=.45]{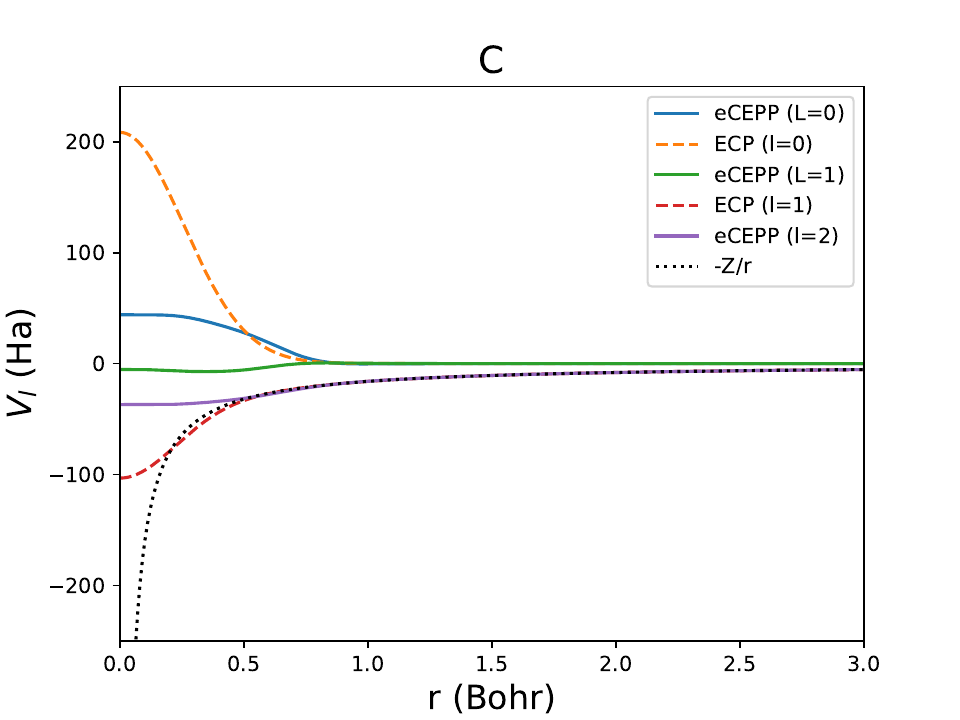}
    \includegraphics[scale=.45]{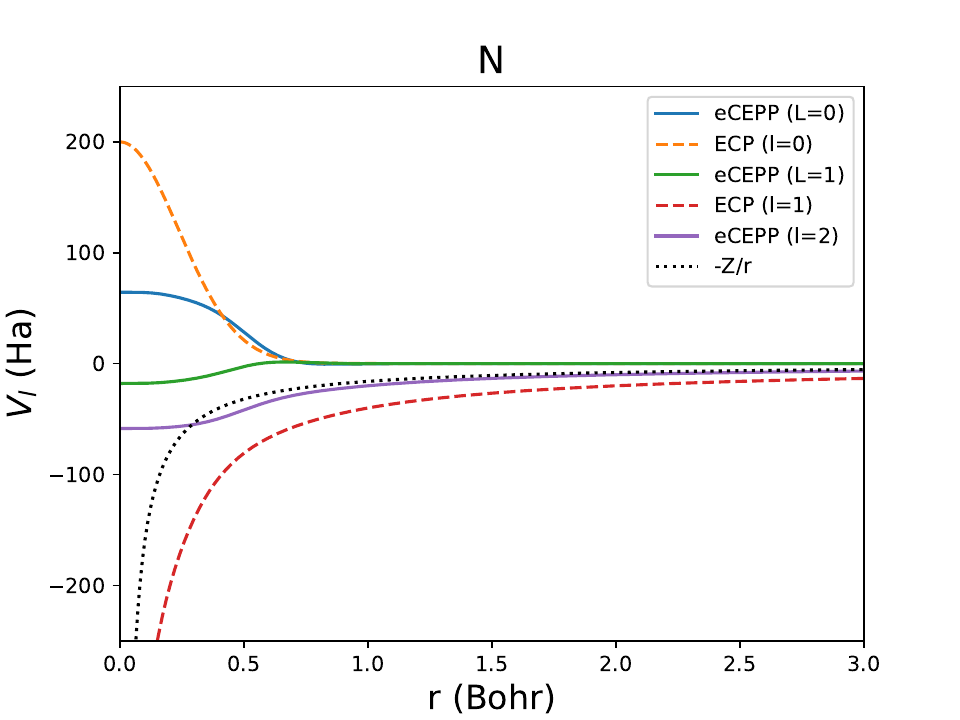}
    \includegraphics[scale=.45]{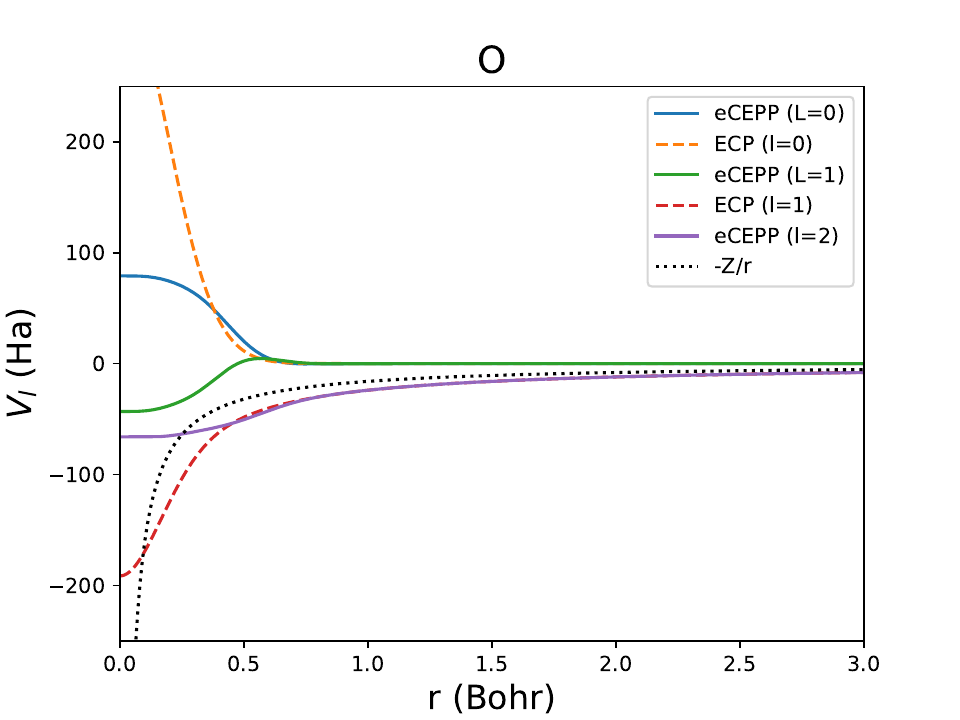}
    \includegraphics[scale=.45]{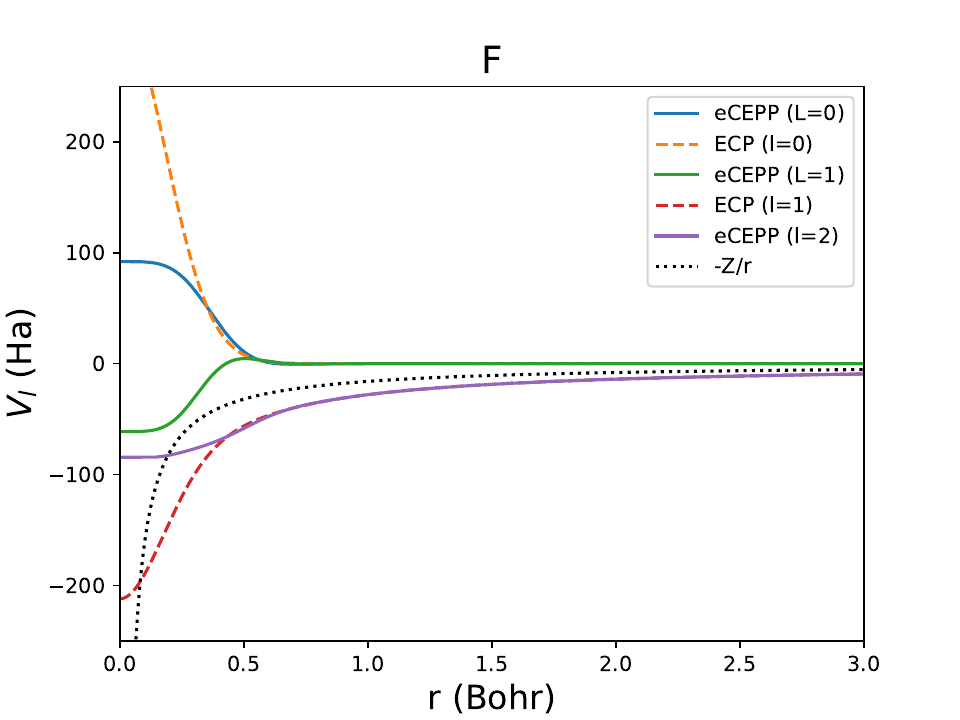}
    \caption{$V_l(r)$ of eCEPPs (solid lines) and ccECPs (dashed lines) of C, N, O, and F. 
        Black dashed lines show the Coulombic potential of the nucleus.}
    \label{fig:ecp visualisation}
\end{figure}

We use two sets of PPs in this work:  energy-consistent 
correlated electron PPs (eCEPPs) by Trail \& Needs \cite{trail2005},
 and correlation-consistent effective core potentials (ccECPs) by Lubos-Mitas \cite{bennett2017} \textit{et al}.
The eCEPPs have $l_\text{max}=2$ for the first-row atoms, with $n=4$ and $m=6$, 
while the ccECPs have $l_\text{max}=1$ and $n=m=1$, leaving only one gaussian term for 
the non-local channels. The different components of ccECPs and eCEPPs for C, N, O, and F are shown 
in Fig. \ref{fig:ecp visualisation}. In the figure, it can be seen that the eCEPPs have much smaller
potential absolute values than the ccECPs.

The action of $\hat{V}_\text{eff}$ on a function $f(\mathbf{r})$ (with or without a 
Jastrow factor) of the position of electron $i$ is:
\begin{equation}
    \hat{V}_\text{eff}(r_i) f(\mathbf{r}_i) = \sum_{l=0}^{l_\text{max}}V_{l}(r_i)  
    \sum_{m=-l}^{l} Y_{lm}(\Omega_{\mathbf{r}_i}) \int_{\left|\mathbf{r}{'}\right|=r_i} 
    d\Omega_{\mathbf{r}{'}}  f(\mathbf{r}{'}) Y_{lm}(\Omega_{\mathbf{r}{'}}).
\end{equation} 
Because $Y_{lm}(f=0,\theta=0)=0$ for $m\neq 0$, we can simplify this expression – 
by choosing the $z$-axis to be along $\mathbf{r}_i$ – to \cite{fahy1990}:
\begin{equation}
    \label{eq:effective potential action on orbital, simplified}
    \hat{V}_{eff}(r_i) f(\mathbf{r}_i) = \sum_{l=0}^{l_{max}}V_{l}(r_i) Y_{l0}
    (\Omega_{\mathbf{r}_i}) \int_{\left|\mathbf{r}{'}\right|=r_i} d\Omega_{\mathbf{r}{'}}  
    f(\mathbf{r}{'}) Y_{l0}(\Omega_{\mathbf{r}{'}}),
\end{equation}
which is the expression we use to estimate the action of the PP on 
the electronic orbitals and the Jastrow factor.

\subsection{Pseudopotentials in the Transcorrelated Hamiltonian\label{section: Pseudopotentials in the Transcorrelated Hamiltonian}}

In transcorrelation theory, the Hamiltonian is supplemented with a Jastrow factor $J$, 
which is a function to describe interparticle correlations. With $N$ electrons and $M$ nuclei
in a system, the Jastrow factor in this work is of Drummond-Towler-Needs (DNT) type \cite{drummond2004}:
\begin{equation}
    \label{eq: j-factor}
    \begin{split}
        J(\{\mathbf{r}_i\},\{\mathbf{R}_I\}) &= \sum_{I=1}^{M} \sum_{i=1}^{N} \chi(\mathbf{r}_{iI}) 
        + \sum_{i<j} u(\left| \mathbf{r}_i-\mathbf{r}_j\right|)  + \sum_{I=1}^{M} \sum_{i<j} f(\mathbf{r}_{iI},\mathbf{r}_{jI},\left| \mathbf{r}_i-\mathbf{r}_j\right|)\\
    &= \sum_{I=1}^{M} \sum_{i>j}J_2(\mathbf{r}_i,\mathbf{r}_j;\mathbf{R}_I),
    \end{split}
\end{equation}
including $1$-body ($\chi, $electron-nucleus), $2$-body ($u$, electron-electron), and $3$-body 
($f$, electron-electron-nucleus) terms. Each of the terms is optimized for interparticle 
distances under a chosen cutoff value, characteristic of the DNT Jastrow form.
Above, $\mathbf{r}_{iI}=\mathbf{r}_i-\mathbf{R}_I$.  In the above, in order to simplify the derviation of the pseudopotential commutator equations \cite{christlmaier2023}, we have included all of the terms involving $\chi, u$ and $f$ into the term $J_2(\mathbf{r}_i,\mathbf{r}_j,\mathbf{R}_I)$.
% to simplify the derivation of pseudopotential commutator equations \cite{christlmaier2023}.
 This approach gives TC contributions to the $2$- and $3$-body terms $\langle pr|qs\rangle$ and $\langle pqr | stu \rangle$
 of the Hamiltonian, including also the $\chi$-term contribution to these terms. We call this approach the "combined" Jastrow treatment. 
 Alternatively, one could treat the $1$- and $2$-body terms in the Jastrow factor separately (which we call the "separate" Jastrow treatment) and
include the contribution of $\chi$ into the $1$-body terms $\langle p|q \rangle$. The separate Jastrow treatment does not remove the contribution of the $\chi$-term from the $2$-body integrals, however, because of the presence of a cross-term (see Supplementary Material). With small  $\chi$ cutoff values the combined and separated treatments yield almost exactly the same coupled cluster total energies for atoms and molecules at equilibrium geometries, with discrepancy of $<.1$mHa.
 However, we found difficulties with the separated Jastrow treatment with PPs whenever the Jastrow cutoff values for the $\chi$ functions exceeded the inter-nuclear bond lengths, and for this reason we chose the combined approach in this study.
 We present the equations for the separated treatment of the $1$- and $2$-body terms in the Jastrow factor 
 in the Supplementary Material.  

The transcorrelated Hamiltonian is obtained from a similarity transformation of the 
original Hamiltonian:
\begin{equation}
    \label{eq:transcorrelated Hamiltonian}
    \hat{H}_{TC} = \exp(-J) \hat{H} \exp(J)= \hat{H} + [\hat{H}, J] + 
    \frac{1}{2!}[[\hat{H}, J], J] + \ldots 
\end{equation}
Any operator component of the Hamiltonian $\hat{O}$ that does not commute 
with the Jastrow factor $J$ will extend the transcorrelated Hamiltonian. For 
single-particle operators, the first two commutators can be expressed as:
\begin{equation}
    \begin{split}
    [\hat{O}, J] =& \sum_{k=1}^{N} \sum_{i<j} 
    \left[ \hat{O}(\mathbf{r}_k), \sum_I^M J_2(\mathbf{r}_i,\mathbf{r}_j; \mathbf{R_I})\right] 
    = \sum_{k=1}^{N} \sum_{i<j} \Pi_{ij}^k(\hat{O}) , \\
    [[\hat{O}, J], J] =& \sum_{k=1}^{N} \sum_{i<j} \sum_{l<m} 
    \left[ \left[ \hat{O}(\mathbf{r}_k),\sum_I^M J_2(\mathbf{r}_i,\mathbf{r}_j; \mathbf{R_I})\right], 
    \sum_J^M J_2(\mathbf{r}_l,\mathbf{r}_m; \mathbf{R_J})\right] 
    = \sum_{k=1}^{N} \sum_{i<j} \sum_{l<m} \Gamma^k_{ijlm}(\hat{O}) .
    \end{split}
\end{equation}
It should be noted that the similarity transformation breaks the variational 
principle, although it does not affect the eigenvalues of the Hamiltonian. 

There are restrictions to the indexing of $\Pi$ and $\Gamma$ terms. 
First, $\Pi^k_{ij}$ is non-zero only when $k=i$ or $k=j$. Second, $\Gamma^k_{ijlm}$ 
is non-zero only when $k=i$ and $i \in (l,m)$ or $k=j$ and $j \in (l,m)$. Hence 
the commutators can be expressed as:
\begin{equation}
    \label{eq:commutators with PI and GAMMA}
    \begin{split}
    \left[ \hat{O},J\right] =& \sum_{i<j} \Pi^i_{ij}(\hat{O}) + \Pi^j_{ij}(\hat{O}) , \\
    [[\hat{O}, J], J] =& \sum_{i>j} \left[ \Gamma^i_{ijij}(\hat{O}) + \Gamma^j_{ijij}(\hat{O}) \right] \\
                       +&\sum_{i>j>m} \left[ \Gamma^i_{ijim}(\hat{O}) + \Gamma^j_{jijm}(\hat{O})  + \Gamma^m_{mimj}(\hat{O}) \right] .
    \end{split}
\end{equation}

In all-electron transcorrelation theory, the only component of the Hamiltonian that 
does not commute with the Jastrow factor is the kinetic energy operator 
\cite{cohen2019}. In this case the similarity transformation of Eq. 
(\ref{eq:transcorrelated Hamiltonian}) terminates exactly at 
the second commutator, and the transcorrelated Hamiltonian is computed with 
the commutators in Eq. (\ref{eq:commutators with PI and GAMMA}), introducing 
$2$- and $3$-body terms in the Hamiltonian.

With PPs, the $V_\text{eff}$ terms do not commute with the Jastrow factor.
In addition, commutator summation in the similarity transformation is not guaranteed to terminate at the 
second commutator. To estimate the effect of the PP on the 
transcorrelated Hamiltonian, we make an approximation of only considering 
the $2$-body terms of the PP commutators, ignoring the sum over 
$i>j>m$ in Eq. (\ref{eq:commutators with PI and GAMMA}). This approximation 
is made under the assumption that the $3$-body interactions of the valence electrons close to the core 
are negligible. Explicit treatment of the $3$-body terms, possibly under the xTC
approximation, is left for future work in case it is needed. Within this 
approximation, the transcorrelated second-quantized Hamiltonian becomes 
\begin{equation}
    \label{eq:transcorrelated Hamiltonian with pseudopotential}
    \begin{split}
    \Hat{H}=\sum_{pq\sigma}h^p_q a^\dagger_{p\sigma}a_{q\sigma} +
     \frac{1}{2}\sum_{pqrs} \left( V^{pq}_{rs} - K^{pq}_{rs}  
     + P^{pq}_{rs} \right) \sum_{\sigma\tau} a^\dagger_{p\sigma}a^\dagger_{q\tau}a_{r\tau}a_{s\sigma} \\
     -\frac{1}{6}\sum_{pqrstu}L^{pqr}_{stu} \sum_{\sigma\tau\lambda} 
     a^\dagger_{p\sigma}a^\dagger_{q\sigma'}a^\dagger_{r\lambda}a_{s\lambda}a_{t\tau}a_{u\sigma} .
    \end{split}
\end{equation}   
The terms $h$ and $V$ are traditional one- and two-body terms of the 
second-quantized Hamiltonian, while the calculation of the transcorrelated 
components $K$ and $L$ arising from the kinetic energy operator has been 
described elsewhere \cite{cohen2019,christlmaier2023}. The pseudopotential 
terms $P$ are computed as:
\begin{equation}
    \label{eq:pseudopotential commutator terms in second quantization}
    \begin{split}
    P^{pq}_{rs} = \Bigg\langle \phi_p\phi_q \left|  \Pi^1_{12}(\hat{H}^\text{PP}_\text{en}) 
    + \Pi^2_{12}(\hat{H}^\text{PP}_\text{en}) + \frac{1}{2}\Gamma^1_{1212}(\hat{H}^\text{PP}_\text{en}) 
    + \frac{1}{2}\Gamma^2_{1212}(\hat{H}^\text{PP}_\text{en}) \right| \phi_r\phi_s \Bigg\rangle
    \end{split}
\end{equation}
with
\begin{equation}
    \label{eq:commutators with PI and GAMMA for pseudopotential}
    \begin{split}
    \Pi^i_{ij}(\hat{H}^\text{PP}_\text{en}) =& \sum_I^M \left[\hat{H}^\text{PP}_\text{en}(\mathbf{r}_i)J_2(\mathbf{r}_i,\mathbf{r}_j; \mathbf{R_I})
    - J_2(\mathbf{r}_i,\mathbf{r}_j; \mathbf{R_I})\hat{H}^\text{PP}_\text{en}(\mathbf{r}_i) \right] , \\
    \Gamma^i_{ijij}(\hat{H}^\text{PP}_\text{en}) =& \hat{H}^\text{PP}_\text{en}(\mathbf{r}_i)\left[\sum_I^M J_2(\mathbf{r}_i,\mathbf{r}_j, \mathbf{R_I}) \right]^2 
    - 2\sum_I^M   J_2(\mathbf{r}_i,\mathbf{r}_j; \mathbf{R_I})\hat{H}^\text{PP}_\text{en}(\mathbf{r}_i)\sum_J^MJ_2(\mathbf{r}_i,\mathbf{r}_j; \mathbf{R_J}) \\
    +& \left[\sum_I^M J_2(\mathbf{r}_i,\mathbf{r}_j; \mathbf{R_I}) \right]^2\hat{H}^\text{PP}_\text{en}(\mathbf{r}_i) .
    \end{split}
\end{equation}
Therefore, in order to evaluate the PP commutators (under the present 
approximation of restricting ccECP corrections to two-body terms)
 in Eq. (\ref{eq:pseudopotential commutator terms in second quantization}), we need to 
 apply Eq. (\ref{eq:effective potential action on orbital, simplified}) to calculate 
 terms of the type $\hat{H}^\text{PP}_\text{en}\phi$, $\hat{H}^\text{PP}_\text{en}J_2\phi$, and 
 $\hat{H}^\text{PP}_\text{en}J_2^2\phi$. This allows us to construct the transcorrelated Hamiltonian 
 in Eq. (\ref{eq:transcorrelated Hamiltonian with pseudopotential}). We have included also 
 higher-order terms in the PP commutators in Eq. (\ref{eq:commutators with PI and GAMMA for pseudopotential})
 in our calculations, under the assumption that the $3$-body terms are negligible.

% If one wants to separate the treatment of the $1$- and $2$-body terms in the Jastrow factor when calculating the 
% pseudopotential commutators, the equations in Eq. (\ref{eq:commutators with PI and GAMMA for pseudopotential}) can be
% rewritten as:
% \begin{equation}
%     \label{eq:commutators with PI and GAMMA for pseudopotential, separated}
%     \begin{split}
%     \Pi^i_{ij}(\hat{H}^{PP}_{en}) =& \hat{H}^{PP}_{en}(\mathbf{r}_i)\chi(\mathbf{r}_i) - \chi(\mathbf{r}_i)\hat{H}^{PP}_{en}(\mathbf{r}_i) \\
%     +& \hat{H}^{PP}_{en}(\mathbf{r}_i)u(\mathbf{r}_i,\mathbf{r}_j) - u(\mathbf{r}_i,\mathbf{r}_j)\hat{H}^{PP}_{en}(\mathbf{r}_i) ,\\
%     \Gamma^i_{ijij}(\hat{H}^{PP}_{en}) =& \hat{H}^{PP}_{en}(\mathbf{R}_i)\chi(\mathbf{r}_i)^2 + 
%     \chi^2(\mathbf{r}_i)\hat{H}^{PP}_{en}(\mathbf{r}_i) - 2\chi(\mathbf{r}_i)\hat{H}^{PP}_{en}(\mathbf{r}_i)\chi(\mathbf{r}_i) \\
%     +& \hat{H}^{PP}_{en}(\mathbf{R}_i)u(\mathbf{r}_i,\mathbf{r}_j)^2 + u(\mathbf{r}_i,\mathbf{r}_j)^2\hat{H}^{PP}_{en}(\mathbf{r}_i) \\
%     -& 2 u(\mathbf{r}_i,\mathbf{r}_j)\hat{H}^{PP}_{en}(\mathbf{r}_i)u(\mathbf{r}_i,\mathbf{r}_j) \\
%     +&2\chi(\mathbf{r}_i)\left( u(\mathbf{r}_i,\mathbf{r}_j)\hat{H}^{PP}_{en}(\mathbf{r}_i)-\hat{H}^{PP}_{en}(\mathbf{r}_i)u(\mathbf{r}_i,\mathbf{r}_j)\right) \\ 
%     +&2\left(\hat{H}^{PP}_{en}(\mathbf{r}_i)u(\mathbf{r}_i,\mathbf{r}_j) - u(\mathbf{r}_i,\mathbf{r}_j)\hat{H}^{PP}_{en}(\mathbf{r}_i)\right)\chi(\mathbf{r}_i) .
% \end{split}
% \end{equation}

\section{Calculations}

\subsection{Evaluation of transcorrelated Hamiltonian with pseudopotentials}

The transcorrelated second-quantized Hamiltonians are calculated
with an in-house code TCHINT \cite{tchint}, that is based on the version used in \cite{cohen2019}.
TCHINT calculates transcorrelated second-quantized Hamiltonians with a Jastrow
factor and the molecular orbitals as inputs. It is parallelized with MPI, uses
the BLAS and LAPACK libraries for matrix operations, and is written in Fortran.
The inclusion of the features necessary for the PP commutator evaluation within
TCHINT has been an important part of this work. 

The elements $P_{rs}^{pq}$ in Eq. (\ref{eq:pseudopotential commutator terms in second quantization})
are obtained with numerical integration over real-space grid points. The operation of $\hat{H}^\text{PP}_\text{en}$
in terms $\Pi$ and $\Gamma$ is evaluated according to Eq. 
(\ref{eq:effective potential action on orbital, simplified}), by discretizing the
spherical integration, so that 
\begin{equation}
    \hat{V}_\text{eff}(r_i) f(\mathbf{r}_i) = \sum_{l=0}^{l_\text{max}}V_{l}(r_i) Y_{l0}
    (\Omega_{\mathbf{r}_i}) \sum_{i=1}^{N_s}  f(\mathbf{r}_i{'}) Y_{l0}(\Omega_{\mathbf{r}_i{'}}) ,
\end{equation}
where $\left|\mathbf{r}_i{'}\right|=\left|\mathbf{r}_i\right|=r_i$ and f is either
$\phi_r(\mathbf{r}_1)$, $J_2(\mathbf{r}_1,\mathbf{r}_2)\phi_r(\mathbf{r}_1)$, or
$J_2(\mathbf{r}_1,\mathbf{r}_2)^2\phi_r(\mathbf{r}_1)$. 
The spherical grid points are chosen as the vertices of an icosahedron. The points 
are obtained by setting them on unit sphere scaled with $r_i$, so that
$[\pm a, \pm b, 0]$, $[\pm b, \pm a, 0]$, and $[0, \pm a, \pm b]$ are the grid points 
$\mathbf{r}_i{'}$ on the unit sphere, with $a=r_i/\sqrt{1+\phi^2}$ and $b=\phi r_i/\sqrt{1+\phi^2}$, 
where $\phi=(1+\sqrt{5})/2$ is the golden ratio. Hence $N_s=12$. This is a Lebedev grid capable 
of exact spherical integration of functions that have up to $l=5$ components.  Tests on denser spherical grids did not change the
results significantly. To mitigate the bias by the orientation of the spherical grid
we applied random rotations to the spherical grid points in each spherical projection.

For a number of grid points $N_\text{ECP}$ under pseudopotential influence,
the numbers of additional Jastrow factor and orbital evaluations due to presence of pseudopotentials
 are $N_\text{g} N_\text{ECP}N_\text{s}$ and $N_\text{ECP}N_\text{orb}N_\text{s}$, respectively, where $N_\text{g}$ is the total number of grid points
and $N_\text{orb}$ is the number of orbitals.
The additional jastrow evaluations thus add a $N_\text{ECP}N_s/N_{g}$ prefactor to the 
computational scaling of the Jastrow evaluations \cite{cohen2019}, which is computationally most 
expensive part of the calculation. 
To mitigate the additional computational cost and load imbalance, we use vectorized Jastrow evaluations and
a load balancing scheme that distributes the grid points evenly among the MPI
processes. 
Orbitals at the spherical grid points are evaluated at the beginning of the calculation, and the 
memory requirements are increased because of ECPs by additional $N_\text{ECP}N_\text{orb}N_\text{s}$
orbital floating point numbers as opposed to the $N_\text{g}N_\text{orb}$ floating point numbers
required for all-electron calculations. 

Table \ref{tab:grid points} shows
 a list of the number of grid points under pseudopotential influence for 
a number of systems studied in this work. For each system, the number of grid points
is roughly 55\% of the total number of grid points.  
\begin{table}[h]
    \scriptsize
    \centering
    \caption{Total and ECP-influenced grid points for each system and pseudopotential.}
    \begin{tabular}{l c c c}
    \toprule
    \textbf{System} & \textbf{Total} & \textbf{eCEPP} & \textbf{ccECP} \\
    \midrule
    C       & 18120 & 10570 & 10268 \\
    N       & 18120 & 10570 & 9966  \\
    O       & 18120 & 10268 & 9362  \\
    F       & 18120 & 9966  & 9060  \\
    \midrule
    N\(_2\) & 36196 & 22792 & 20810 \\
    O\(_2\) & 36198 & 21380 & 19058 \\
    F\(_2\) & 36202 & 20316 & 18324 \\
    CN      & 36196 & 22445 & 21045 \\
    CO      & 36196 & 22022 & 20313 \\
    CF      & 36198 & 21299 & 19767 \\
    \midrule
    H\(_2\)O & 33484 & 20066 & 9612  \\
    CO\(_2\) & 54056 & 33384 & 30250 \\
    \bottomrule
    \end{tabular}
    \label{tab:grid points}
\end{table}

The treatment of the $3$-body terms when constructing the transcorrelated Hamiltonian is done under the xTC approximation \cite{christlmaier2023}. In this 
approximation, the last term of Eq. \ref{eq:transcorrelated Hamiltonian 
with pseudopotential} containing $L$ is reorganized within the generalized 
normal ordering scheme. This leads to modifications in the $1$-,$2$-, and 
$3$-body terms of the transcorrelated Hamiltonian. The contributions of the $3$-body terms under 
the generalized normal ordering to the $1$- and $2$-body terms are 
evaluated by contracting the $3$-body terms with the reduced $1$-body density matrix of the 
Hartree-Fock wave function (with the exception of using FCI 
density matrix in the N\textsubscript{2} dissociation curve calculations. 
The remaining  3-body terms are neglected in xTC.

\subsection{General computational details}

In this work we study the use of pseudopotentials with the transcorrelated 
Hamiltonian in atoms Be, B, C, N, O, and F, as well as their $+1$ ions. We 
also study the total and atomization energies of molecules CN, CO, CF, N$_2$, 
O$_2$, F$_2$, H$_2$O and CO$_2$.  We use the 
aug-cc-pVD/T/QZ basis sets (AVXZ, with X=D,T,Q), optimized individually for each 
pseudopotential\cite{trail2005,bennett2017}. Ionization energies with ccECPs are evaluated with non-augmented  
cc-pVD/T/QZ basis sets (PVXZ, with X=D,T,Q). 
.

The geometries of the molecules are taken from those in the HEAT database 
\cite{tajti2004}. The Hartree-Fock (HF) calculations are done with the PYSCF 
code \cite{sun2020}. The Jastrow factors used are of Drummond-Towler-Needs
type \cite{drummond2004} and are optimised with respect to the variance
of the Hartree-Fock wave function using the VMC method of the CASINO package 
\cite{needs2020}. The Jastrow factors are optimized separately for each
system, basis set, and pseudopotential combination. The cutoffs used for the 
$u$, $\chi$, and $f$ terms were $4.5$, $4$, and $4$, respectively. We provide the optimized
Jastrow factors in the Supplementary Material.

The atomization energies are calculated both with and without transcorrelation 
with the coupled cluster using singles, doubles 
and perturbative and full triples (CCSD(T) and CCSDT) using the ElemCo.jl
package \cite{elemcojl}. If transcorrelation is used, we add a prefix 
xTC- to the method name. The similarity transformation of the Hamiltonian using the 
Jastrow factor leads to a non-Hermitian Hamiltonian with a non-diagonal Fock matrix. 
Consequently, standard non-iterative perturbative methods for CCSD(T) are not directly 
applicable to the transcorrelated Hamiltonian. Thus for transcorrelated CCSD(T) we do the
calculations with a pseudocanonical $\Lambda$CCSD(T) approach using bi-orthogonal orbitals 
\cite{kats2024}. To simplify notation we will call it xTC-CCSD(T).

The atomization energies are also calculated
with CCSD(T)-F12  using the Molpro package \cite{werner2012molpro,werner2020,MOLPRO} for comparison.
Both all-electron and PP F12 results are 
calculated to assess the effect of the PPs on the results.
In all-electron calculations we use the standard AVXZ family of basis sets
with AVXZ-MP2Fit and VXZ-JKFit auxiliary basis sets \cite{peterson2008}. 
In PP calculations we use the 
augmented basis sets fitted for PPs \cite{trail2005,bennett2017}. 
As the auxiliary basis sets with PPs
we use mp2-fitted QZVPP/MP2Fit and TZVPP/MP2Fit basis sets.

For the dissociation curves, we use the full configuration interaction quantum Monte 
Carlo method (FCIQMC) \cite{cleland2010} with the NECI package \cite{guther2020} both 
with and without transcorrelated Hamiltonians. The initiator
approximation \cite{cleland2010} with an initiator threshold of $3$ is 
used in the FCIQMC calculations. The walker number for each calculation 
was increased by a factor of $5$ until the energy was converged
to within $1$ mHa. We compare the results with MRCI-F12 calculations, 
done with the Molpro package \cite{werner2012molpro,werner2020,MOLPRO}.
The Davidson correction \cite{davidson1977} is used in the MRCI-F12 calculations.

\subsection{Notation}

All of the calculations presented in the following sections – with the 
exception of some of the F12-calculations – are done with PPs. 
To estimate the effect of evaluating the PP commutators
we do the transcorrelated calculations without the PP 
commutators, and with varying level of commutator evaluations. 
We refer to these calculations as xTC-\{method\}(PP-n), with 
n indicating the level of commutator evaluation ($0-4$ commutator 
evaluations in this work) and method indicating the method used
(CCSD(T), CCSDT, FCIQMC).

When presenting the total energies of atoms, ions, and
molecules with PPs, we show the results relative to an 
energy $E_\text{CBS}^\text{TZ-QZ}$, evaluated as the sum of Hartree-Fock energy in the
PVQZ (ccECPs) or AVQZ (eCEPPs) basis set, $E_\text{HF}^\text{QZ}$, and 
the estimate of the complete basis set limit (CBS) of CCSD(T) correlation energy.
This estimate is obtained from the PVTZ and PVQZ (ECP) or
AVTZ and AVQZ (eCEPP) correlation 
energies $E^\text{Corr}_\text{TZ}(\text{CCSD(T)})$ and $E^\text{Corr}_\text{QZ}(\text{CCSD(T)})$
with a linear extrapolation, so that 
\begin{equation}
    %E_{CBS}^{TZ-QZ}(\text{CCSD(T)}) = \frac{3^3E_{TZ}-4^3E_{QZ}}{3^3-4^3} 
    E_\text{CBS}^\text{TZ-QZ}(\text{CCSD(T)}) = E_\text{HF}^\text{QZ} + \frac{3^3E^\text{Corr}_\text{TZ}(\text{CCSD(T)})-4^3E^\text{Corr}_\text{QZ}((\text{CCSD(T)}))}{3^3-4^3}.
    \label{eq:extrapolated energy1}
\end{equation}
This estimation of the complete basis set limit energy with respect to the
triple- and quadruple-zeta basis sets is not meant to serve as a
benchmark, but rather as a reference point for the PP energies for 
easier comparison.

\section{Results}

\subsection{Variances in VMC optimization}

Table \ref{tab:variance_data} shows the variance of the reference energy in Hartrees 
for the first row elements Be-F, and for a set of first row molecules, obtained 
from eCEPP, ccECP, and all-electron (AE) calculations. For the results, 
we sampled the reference Hartree-Fock wave function with the Metropolis algorithm and
evaluated an estimate of the variance of the obtained configurations together with the optimized
Jastrow factor. The percentages after the PP
variances show the ratio of PP values against the all-electron variances.
The variance is significantly
reduced when using PPs as compared to AE calculations. The variance reduction is greater for 
the heavier atoms. For the atoms, the variance reduction seems to obey roughly $1/N_v$ dependence,
where $N_v$ is the number of valence electrons.

The non-local PP cutoff is generally lower for ccECPs than for eCEPPs 
(see Table \ref{tab:cutoff_radius}). With the ccECPs and eCEPPs we got almost identical
variances, which hints that the variance is stable against the cutoff 
radius. 

\begin{table}[H]
    \centering    
    \caption{
        \label{tab:variance_data}
        Variance data for atoms and molecules using AE, eCEPP, and ccECP methods.
        Units are in Hartrees. The percentages after the PP
        variances show the ratio of PP values against the all-electron variances.
    }
    \begin{tabular}{lccc}
        \toprule
        \textbf{Atom/Molecule} & \textbf{AE} & \textbf{eCEPP} & \textbf{ccECP} \\ 
        \midrule
        \textbf{Be}    & 0.0586  & 0.0146 (25\%) & 0.0180 (31\%) \\ 
        \textbf{B}     & 0.224   & 0.045 (20\%)  & 0.049 (22\%)  \\ 
        \textbf{C}     & 0.510   & 0.085 (17\%)  & 0.082 (16\%)  \\ 
        \textbf{N}     & 1.110   & 0.144 (13\%)  & 0.135 (12\%)  \\ 
        \textbf{O}     & 2.290   & 0.254 (11\%)  & 0.243 (11\%)  \\ 
        \textbf{F}     & 4.300   & 0.400 (9\%)   & 0.364 (8\%)   \\ 
        \textbf{N\(_2\)} & 2.067 & 0.4441 (21\%) & 0.4313 (21\%) \\ 
        \textbf{O\(_2\)} & 3.734 & 0.7133 (19\%) & 0.6741 (18\%) \\ 
        \textbf{F\(_2\)} & 6.428 & 0.9423 (15\%) & 0.8828 (14\%) \\ 
        \textbf{CN}    & 1.482   & 0.3353 (23\%) & 0.3188 (22\%) \\ 
        \textbf{CF}    & 3.753   & 0.5674 (15\%) & 0.5336 (14\%) \\ 
        \textbf{CO}    & 2.483   & 0.4602 (19\%) & 0.4321 (17\%) \\ 
        \textbf{H\(_2\)O} & 1.763 & 0.3100 (18\%) & 0.2917 (17\%) \\ 
        \textbf{CO\(_2\)} & 4.236 & 0.7683 (18\%) & 0.7212 (17\%) \\ 
        \bottomrule
    \end{tabular}
\end{table}

\begin{table}[H]
    \centering
    \caption{
        \label{tab:cutoff_radius}
        Non-local cutoff radius values (Bohr) for various atoms, in atomic units, 
        for eCEPPs and ccECPs. The cutoff is defined as the radius above which the 
        PP radial functions are less than $10^{-6}$ Ha.
    }
    \begin{tabular}{lcccccc}
        \toprule
        \textbf{Atom}   & \textbf{Be}  & \textbf{B}   & \textbf{C}   & \textbf{N}   & \textbf{O}   & \textbf{F}   \\ 
        \midrule
        \textbf{eCEPP}  & 2.735610     & 2.066929     & 1.603693     & 1.608516     & 1.427703     & 1.324644     \\ 
        \textbf{ccECP}    & 2.405539     & 1.897133     & 1.430486     & 1.341815     & 1.088303     & 1.039012     \\ 
        \bottomrule
    \end{tabular}
\end{table}

Because the variance of the VMC energy is smaller with PPs,
one can optimize the Jastrow factor with fewer Monte Carlo samples and 
hence less computational cost. 

\subsection{Analysis of the transcorrelated integrals}

\begin{figure*}
    \includegraphics[scale=.55]{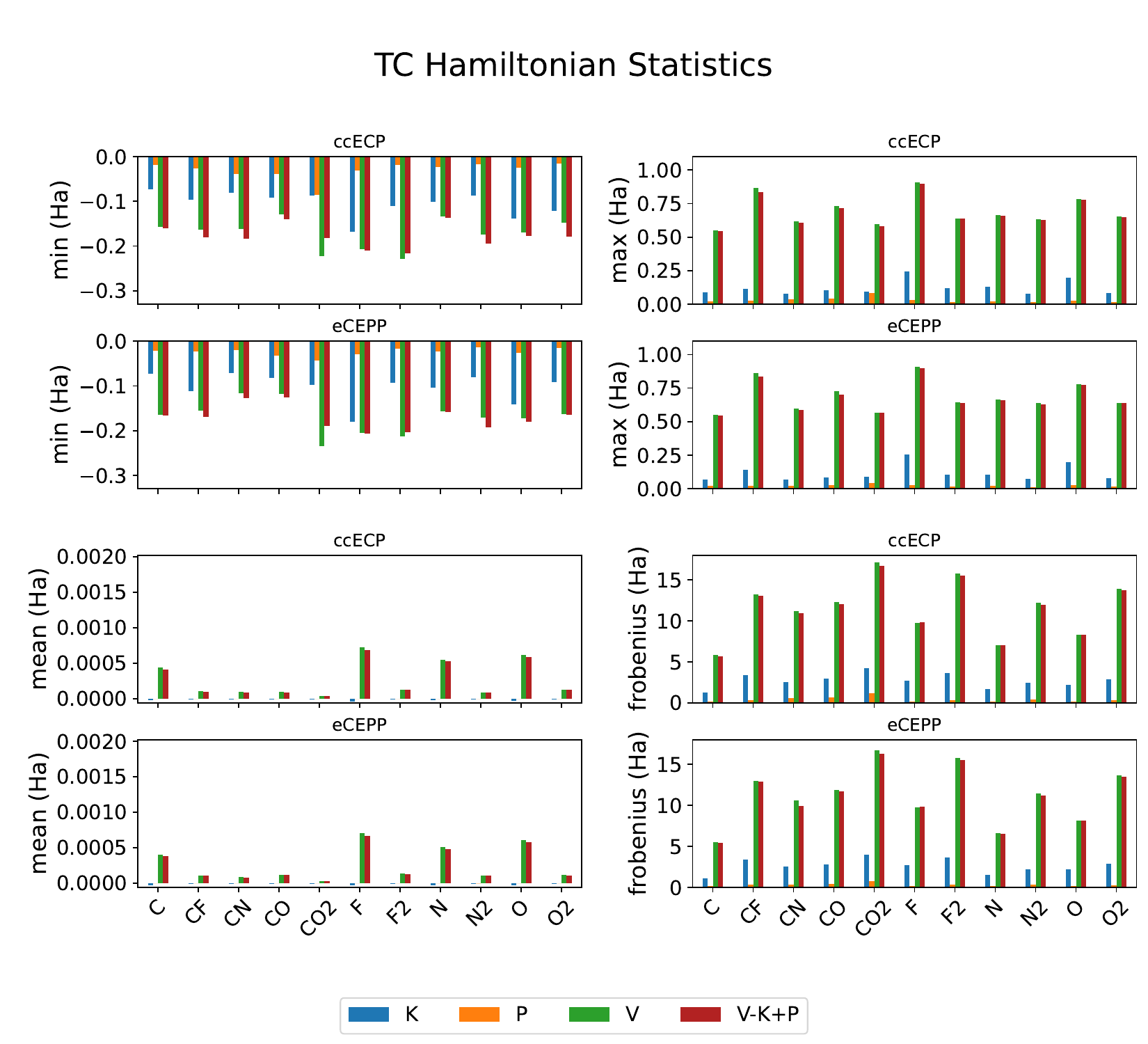}
    \caption{\label{fig:pytchint_statistics} Parameters of the off-diagonal values of $V$ (green), $K$ (blue), $P$ (orange),
    and $V-K+P$ (red)
    tensors of Eq. \ref{eq:transcorrelated Hamiltonian with pseudopotential} for different 
    systems. The minima (upper left), maximas (upper right), mean values (lower left),
    and Frobenius norms (lower right) of the tensors are shown in Hartrees. For each stochastic parameter,
    we show values for each system for both ccECPs and eCEPPs.}
\end{figure*}

For the systems studied in this work, we have investigated statistical parameters of the off-diagonal values of 
 $V$, $K$, and
$P$ tensors, as well as the full TC Hamiltonian. The minima, maximas,
mean values, and Frobenius norms of the tensors are shown in Fig. \ref{fig:pytchint_statistics}
for both ccECPs and eCEPPs. The values are shown in Hartrees. Frobenius norm is defined as 
$F=\sqrt{\sum_{ij}A_{ij}^2}$ for a matrix $A$.

The data shows that the largest values and overall weight of the off-diagonals are in the $V$ tensor, i.e., 
the non-transcorrelated Hamiltonian, with the full TC Hamiltonian having slightly smaller values and overall 
weight (the Frobenius norm) than $V$. The $K$ tensor of the kinetic energy operator commutators 
has much smaller weight compared to the full Hamiltonian, and the statistical parameters of the $P$
tensor show that the pseudopotential commutators introduce only a slight correction to the TC 
Hamiltonian. 

This data shows that pseudopotential commutators introduce generally small but 
non-negligible contributions.

\subsection{Atoms Be-F}

\subsubsection{Analysis of the degree of PP commutators}

\begin{figure}[H]%\hspace*{-6cm}
    \centering
    \begin{subfigure}{0.65\textwidth}
        %\centering
        \caption{eCEPPs:}
        \hspace*{-3cm}\includegraphics[scale=0.5]{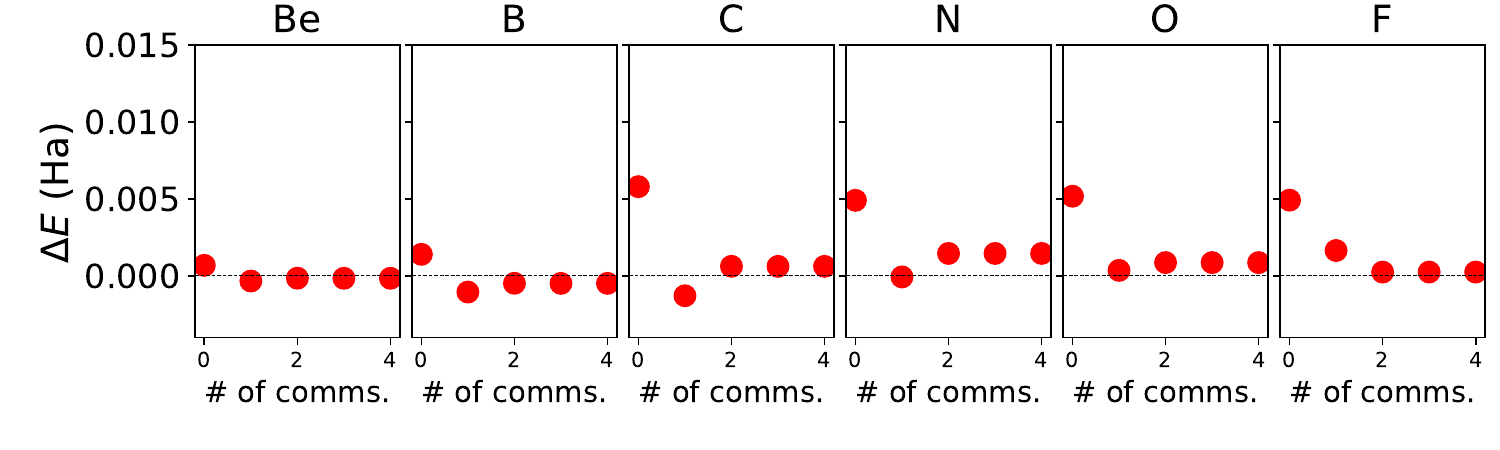}
        \hspace*{-3cm}\includegraphics[scale=0.5]{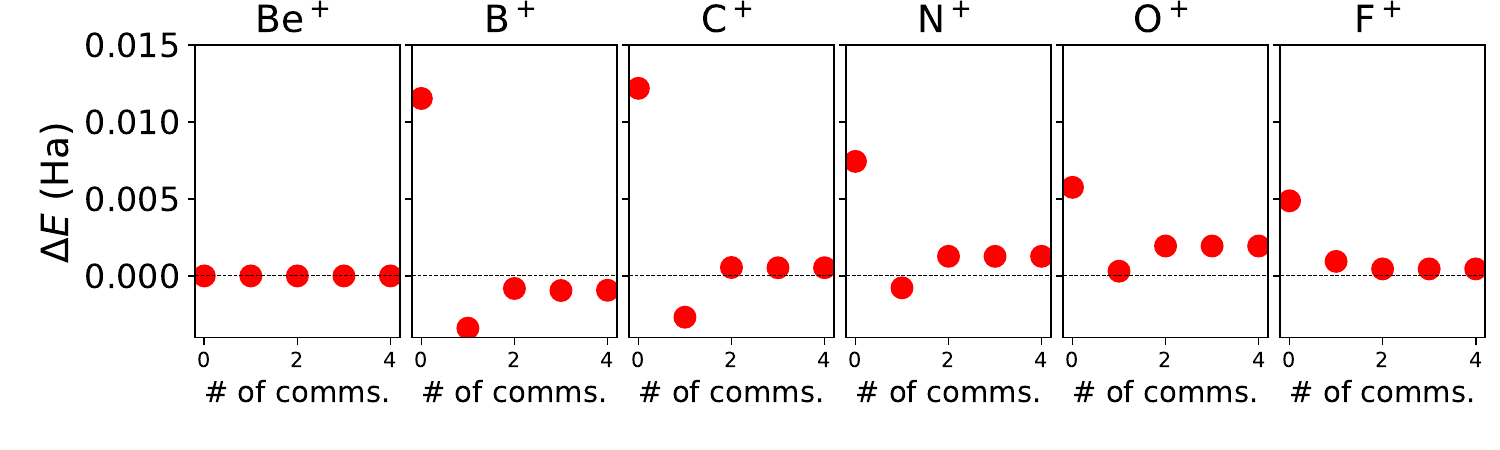}
    \end{subfigure}
    \begin{subfigure}{0.65\textwidth}\hspace*{-126cm}
        %\centering
        \caption{ccECPs:} 
        \hspace*{-3cm}\includegraphics[scale=0.5]{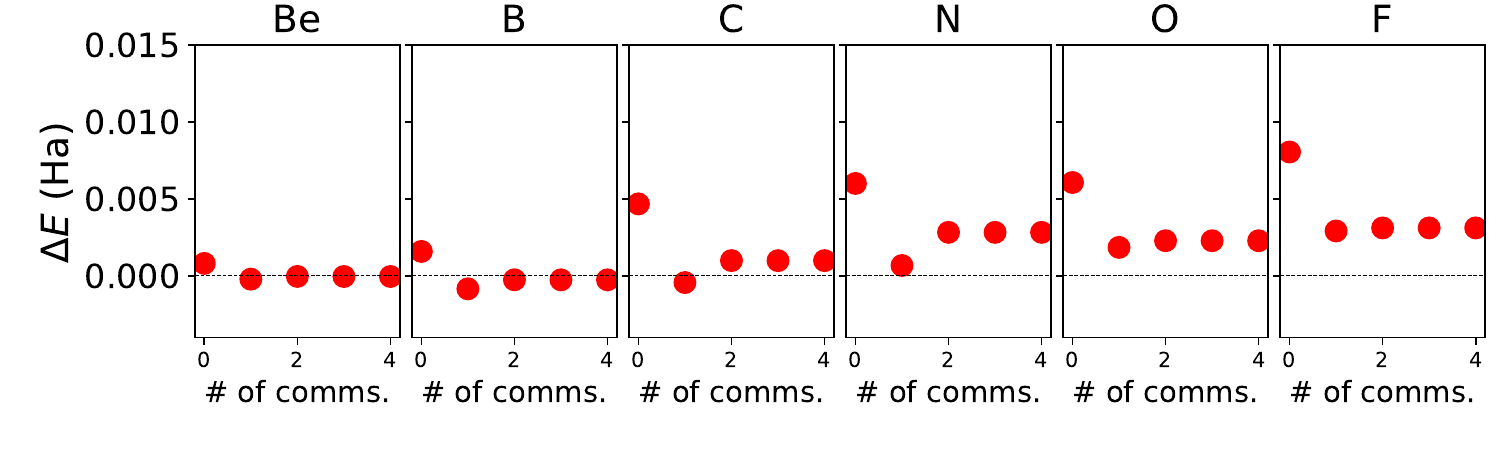}
        \hspace*{-3cm}\includegraphics[scale=0.5]{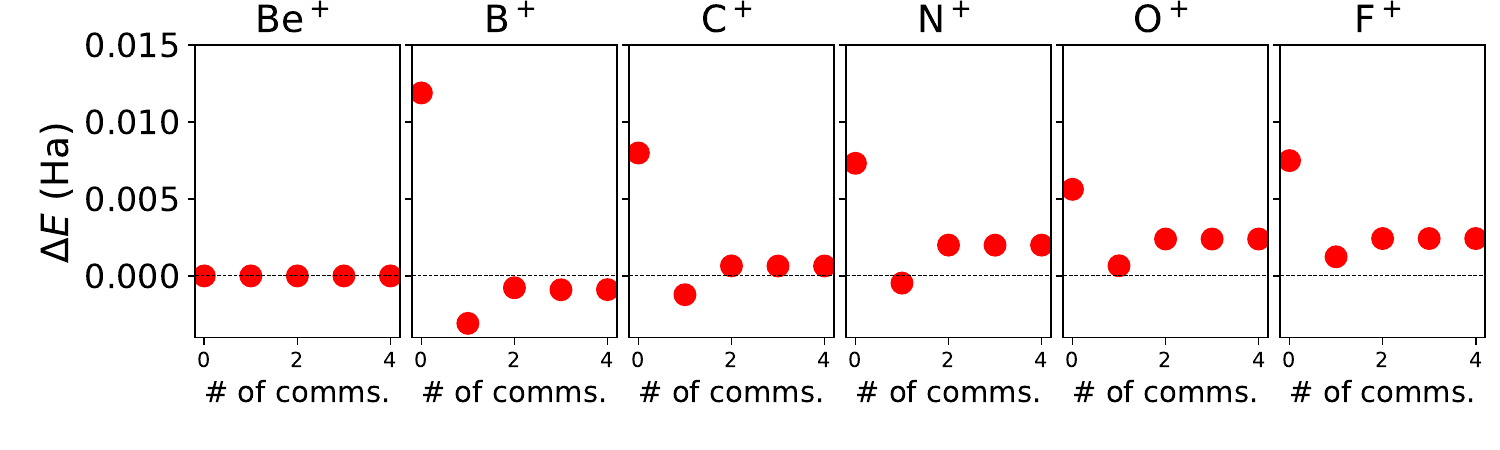}
    \end{subfigure}
    \caption{
    xTC-CCSD(T)(PP-$n$)  energies with eCEPPs (figure (a), AVQZ basis) 
    and ccECPs (figure (b), PVQZ basis) for the neutral and ionised states 
             of the first row elements considered, shown as a function 
             of the degree of PP commutators $n$. Results with 
             $n=0,1,2,3,4$ are shown. Results are presented 
             relative to CBS estimate as $\Delta E=E(\text{xTC-CCSD(T)(PP-}n))-E_\text{CBS}^\text{TZ-QZ}(\text{CCSD(T)})$,
             see Eq. (\ref{eq:extrapolated energy1})}
    \label{fig: pseudopotential commutator degree}
\end{figure}

In Fig. \ref{fig: pseudopotential commutator degree} we show the xTC-CCSD(T)(PP-$n$) 
total energies of the first row elements Be-F, along with their 
ionised states, as a function of $n$, the degree of PP commutators 
evaluated. Results with both eCEPPs (a) and ccECPs (b) are shown. The 
results are calculated with the quadruple-zeta basis set. The zero in the 
y-axis refers to $E_\text{CBS}^\text{TZ-QZ}(\text{CCSD(T)})$ (see Eq. (\ref{eq:extrapolated energy1})).

The Be cation with PPs has only one electron, and hence there is no correlation energy. 
With other atoms and ions, the general trend is that the
xTC-CCSD(T)(PP-$1$)  are smaller than the xTC-CCSD(T)(PP-$0$) energies, and the
xTC-CCSD(T)(PP-$2$) energies increase slightly from the xTC-CCSD(T)(PP-$1$) energies.
The energy converges with the $2$nd order commutator evaluation for all of the systems,
except for the B ion, where the $3$rd order commutator evaluation still decreases the energy,
although the difference is small.

\begin{figure}[H]
    \centering
    \begin{subfigure}{0.65\textwidth}
        %\centering
        \caption{\textbf{1st ionisation energies, eCEPPs:}} 
        \hspace*{-2cm}\includegraphics[scale=.5]{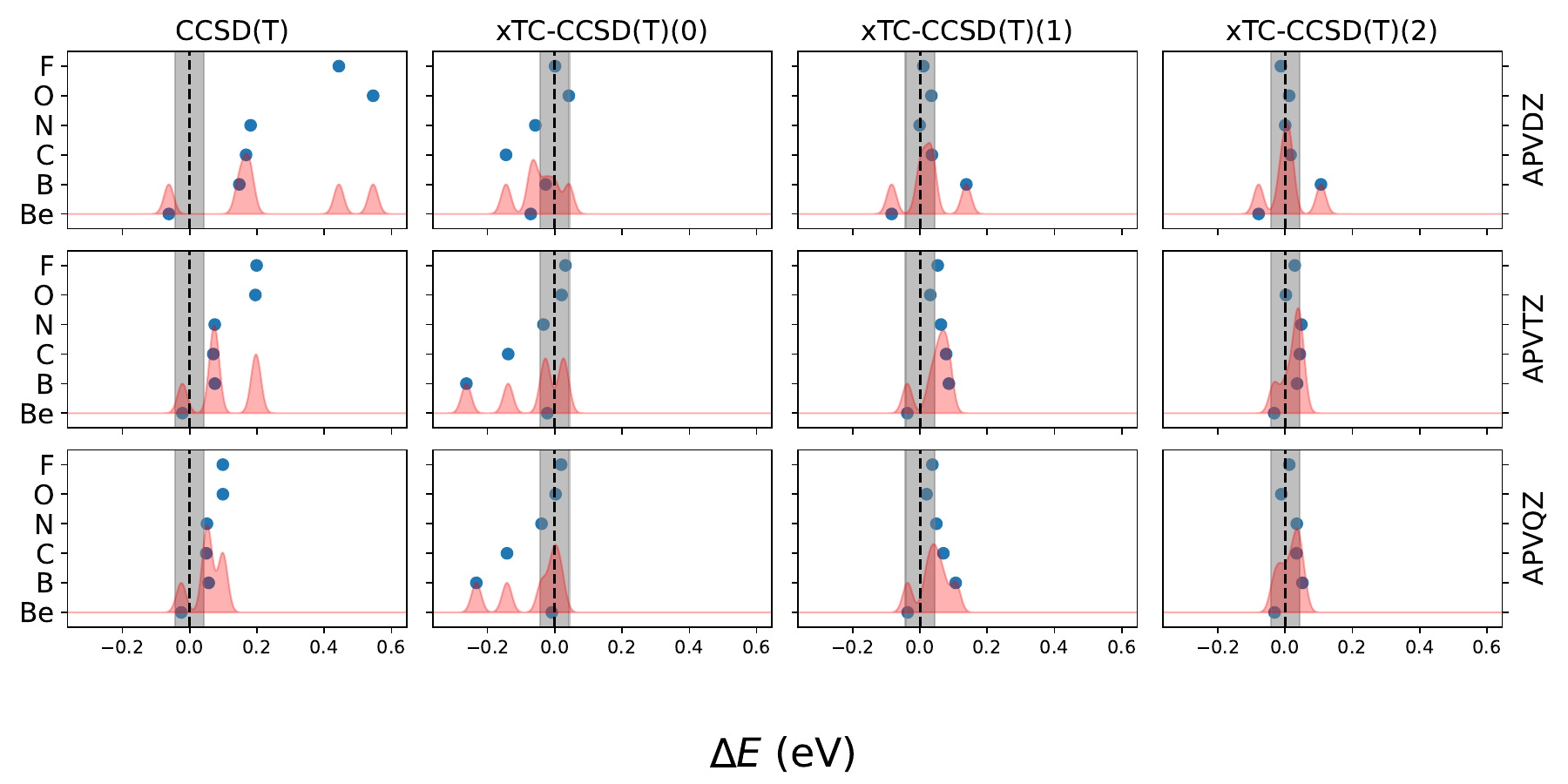}
    \end{subfigure}
    \begin{subfigure}{0.65\textwidth}%\hspace*{-126cm}
        %\centering
        \vspace*{1cm}\caption{\textbf{1st ionisation energies, , ccECPs:}} 
        \hspace*{-2cm}\includegraphics[scale=.5]{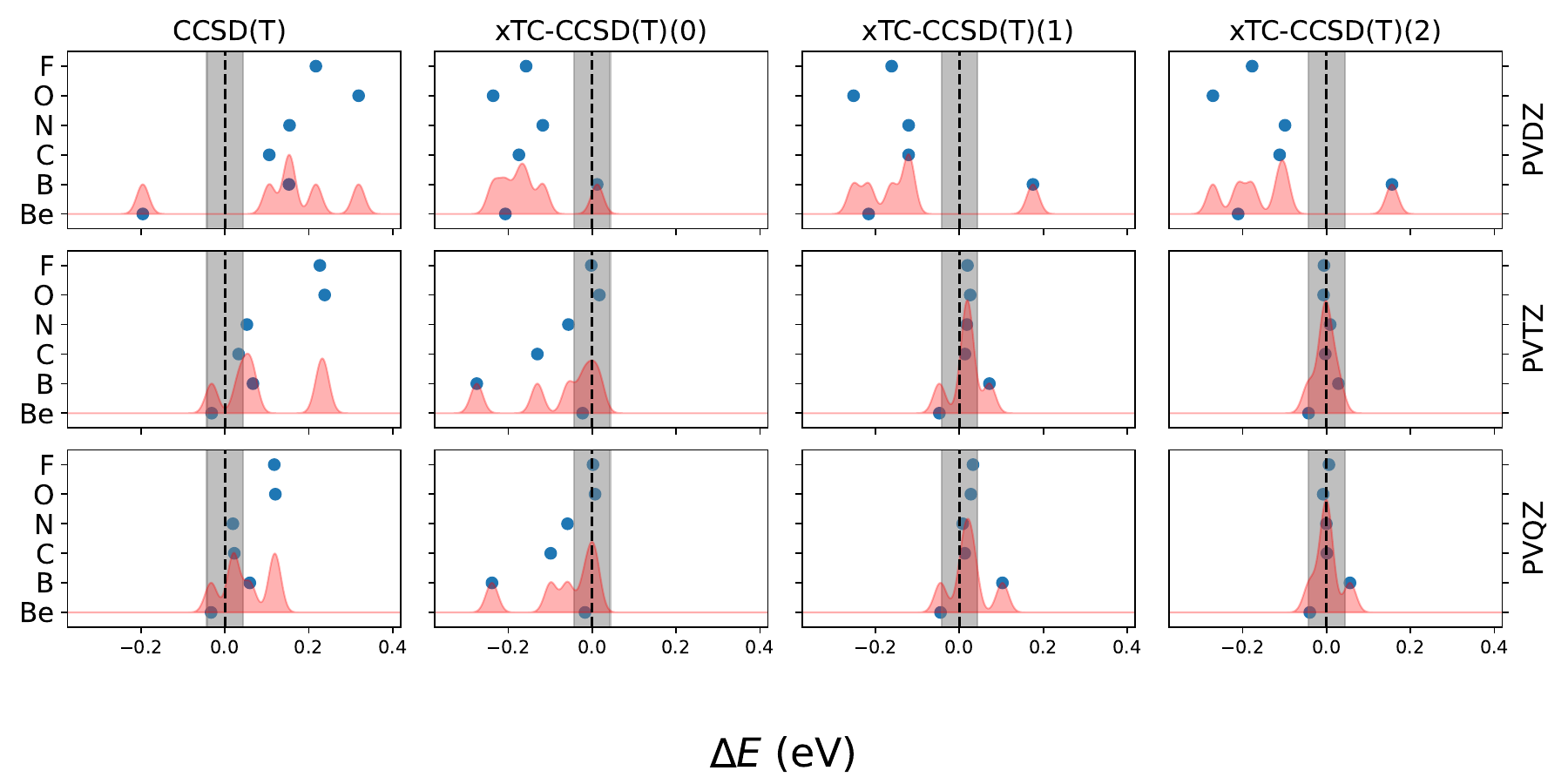}
    \end{subfigure}

    \caption{ Ionisation energies (IPs) $E_\text{i}=E_\text{atom}-E_\text{ion}$ for the 
    first row elements, using (a) eCEPPs and (b) ccECPs. The energies are  presented as the discrepancy 
    with the experimental ionisation energies \cite{chakravorty1993}, so that the presented energies 
    are $\Delta E = E_\text{method}^\text{basis} - E_\text{exp}$. The X-axis shows the xTC-CCSD(T)(PP-$n$) ionisation energies
    with $n=0,1,2$ (in 2nd, 3rd, and 4th column, respectively). The 1st column 
    shows CCSD(T) ionisation energies. The basis sets used are the AVXZ series for eCEPPs and the PVXZ series 
    for ccECPs, where X is either D, T, or Q in the 1st, 2nd, and 3rd row, respectively. The grey shaded 
    region denotes chemical accuracy. The red shading denotes a sum of gaussians, centered at each 
    data point, with a width set such that equidistant gaussians in the presented scale would overlap 
    at $95$ \% confidence.
    \label{fig: ionisation energies}}

\end{figure}

Figure \ref{fig: ionisation energies} shows the ionisation energies of the first row elements Be-F,
obtained with CCSD(T), xTC-CCSD(T)(PP-$0$), xTC-CCSD(T)(PP-$1$), and xTC-CCSD(T)(PP-$2$) using both
eCEPPs (a) and ccECPs (b). The results are shown as a function of the basis set. The ionisation energies
are shown against experimental values \cite{chakravorty1993}.

CCSD(T) ionisation energies reach chemical accuracy  ($1.6$mHa, $0.04$eV)
for Be, C, and N with ccECPs and PVQZ basis set. With eCEPPs and 
AVQZ basis only Be ionisation energy with CCSD(T) is chemically
accurate, but B, C, and N are close to chemical accuracy.

With xTC-CCSD(T)(PP-$2$) we reach chemical
accuracy already with the AVTZ basis set for all of the atoms, with both PPs. 
The evaluation of the PP commutators is seen to be important,
as with PP commutator degrees $n<2$ the xTC-CCSD(T)(PP-$n$) 
ionisation energies are generally worse.

Table \ref{tab:ionisation_errors} shows the mean absolute and root mean square errors
(MAE and RMS) of the ionisation energies, evaluated against experimental values, of the first row 
elements Be-F, obtained with CCSD(T) and xTC-CCSD(T)(PP-$n$) methods and $n=0-4$. The errors are averaged
over the 1st-row atoms studied, and are shown separately for each basis set. Results are shown for both
eCEPPs and ccECPs. 

Table \ref{tab:ionisation_errors} shows that CCSD(T),xTC-CCSD(T)(PP-$0$), and xTC-CCSD(T)(PP-$1$) methods do
not reach chemical accuracy with respect to MAE and RMS for the ionisation energies of the first row elements 
with any of the basis sets. xTC-CCSD(T)(PP-$0$) is even worse in accuracy than CCSD(T) in quadruple-zeta basis sets.
However, xTC-CCSD(T)(PP-$2$) reaches chemical accuracy 
for both the MAE and MSE with both of the PPs at triple and quadruple-zeta basis sets.

% \begin{figure}[H]
%     \centering
%     \begin{subfigure}{0.65\textwidth}
%         %\centering
%         \caption{eCEPPs:}
%         \hspace*{-5cm}\includegraphics[scale=0.5]{first_row_elemco_basis_set_convergence_atom.pdf}
%         \hspace*{-5cm}\includegraphics[scale=0.5]{first_row_elemco_basis_set_convergence_ion.pdf}
%     \end{subfigure}
%     \begin{subfigure}{0.65\textwidth}\hspace*{-126cm}
%         %\centering
%         \caption{ccECPs:}
%         \hspace*{-5cm}\includegraphics[scale=0.5]{first_row_lm_elemco_basis_set_convergence_atom.pdf}
%         \hspace*{-5cm}\includegraphics[scale=0.5]{first_row_lm_elemco_basis_set_convergence_ion.pdf}
%     \end{subfigure}
%     \caption{Convergence of CCSD(T) (Green), xTC-CCSD(T)(PP-$0$) (Yellow), 
%     and xTC-CCSD(T)(PP-$2$) (Violet) energies for the first row elements, 
%     shown as a function of the basis set. The presented energies are
%     relative to the extrapolated CBS limit as  
%     $\Delta E = E - E_{CBS}^{TZ-QZ}(\text{CCSD(T)})$, see Eq. (\ref{eq:extrapolated energy1}).} 
%     \label{fig: total atom energies as a function of basis set}
% \end{figure}

It is interesting to note that xTC methods in double-zeta basis set are much better with eCEPPs than ccECPs,
while the ccECPs are better with higher-order basis sets when $n>0$. This 
phenomenon is not seen with standard CCSD(T), where the accuracy is similar with both
PPs at the same basis set cardinal number. Another feature visible in this table is that the best results
with xTC-CCSD(T) and ccECPs are obtained with the PVTZ basis set, and not with
the PVQZ basis set, again when $n>0$. This is not true with eCEPPs or with CCSD(T). With 
eCEPPs and xTC-CCSD(T)(PP-$2$) the results are converged in AVTZ basis.

When comparing the results with 2 and 3 commutators evaluated, the MAE and MSE are
within $1$ mHa. The 4th commutator produces practically identical results to the
3rd commutator. 

\begin{table}[h!]
    \centering    
    \caption{\label{tab:ionisation_errors}Mean absolute and root mean square errors (MAE and MSE) 
    against the experimental ionisation energies of CCSD(T) and xTC-CCSD(T)(PP-$n$) methods with $n=0-4$. 
    Results are obtained with eCEPPs and ccECPs across different basis sets. The errors are in eV}
    \begin{tabular}{lccccccc}

        \toprule
        \textbf{\# of comms.} & \textbf{Error} & \multicolumn{3}{c}{\textbf{eCEPP (eV)}} & \multicolumn{3}{c}{\textbf{ccECP (eV)}} \\
        \cmidrule(lr){3-5} \cmidrule(lr){6-8}
                      &                & \textbf{AVDZ} & \textbf{AVTZ} & \textbf{AVQZ} & \textbf{PVDZ} & \textbf{PVTZ} & \textbf{PVQZ} \\
        \midrule
        \multirow{2}{*}{\textbf{CCSD(T)}} & MAE & 0.2520 & 0.1001 & 0.0574 & 0.1846 & 0.1020 & 0.0560 \\
                                         & RMS  & 0.3047 & 0.1185 & 0.0624 & 0.1966 & 0.1330 & 0.0691 \\
        \midrule
        \multirow{2}{*}{\textbf{xTC-CCSD(T)(PP-$0$)}} & MAE  & 0.0630 & 0.0833 & 0.0737 & 0.1565 & 0.0899 & 0.0756 \\
                                    & RMS  & 0.0754 & 0.1234 & 0.1136 & 0.1724 & 0.1283 & 0.1099 \\
        \midrule
        \multirow{2}{*}{\textbf{xTC-CCSD(T)(PP-$1$)}} & MAE  & 0.0530 & 0.0520 & 0.0470 & 0.1803 & 0.0277 & 0.0338 \\
                                    & RMS  & 0.0694 & 0.0566 & 0.0568 & 0.1859 & 0.0371 & 0.0476 \\
        \midrule
        \multirow{2}{*}{\textbf{xTC-CCSD(T)(PP-$2$)}} & MAE  & 0.0435 & 0.0259 & 0.0266 & 0.1769 & 0.0191 & 0.0225 \\
                                    & RMS  & 0.0568 & 0.0303 & 0.0302 & 0.1858 & 0.0243 & 0.0296 \\
        \midrule
        \multirow{2}{*}{\textbf{xTC-CCSD(T)(PP-$3$)}} & MAE  & 0.0439 & 0.0267 & 0.0273 & 0.1772 & 0.0197 & 0.0231 \\
                                    & RMS  & 0.0574 & 0.0313 & 0.0314 & 0.1860 & 0.0251 & 0.0306 \\
        \midrule
        \multirow{2}{*}{\textbf{xTC-CCSD(T)(PP-$4$)}} & MAE  & 0.0439 & 0.0267 & 0.0272 & 0.1772 & 0.0197 & 0.0230 \\
                                    & RMS  & 0.0573 & 0.0312 & 0.0312 & 0.1860 & 0.0251 & 0.0306 \\
        \bottomrule
        \end{tabular}
\end{table}

To conclude this section, we have shown that it is necessary to include at least the second commutator, i.e. the PP-$2$ approximation, to achieve chemical 
accuracy in the ionisation potentials of the first-row atoms with the xTC-CCSD(T)-PP-$n$ method. In other words, the first two non-zero commutators arising from the non-local pseudopotentials with the Jastrow factors are critically important to maintain reliability in the TC method with ECPs. Going to higher order commutators do not significantly change the results and hence can be disregarded. In further work we employ the PP-2 approximation.

\clearpage

\subsection{Molecules}

\subsubsection{Total transcorrelated energies}

\begin{figure}[H]\centering
    \begin{subfigure}{0.6\textwidth}
        \centering
        \caption{eCEPPs:} 
        \hspace*{-5cm}
    \includegraphics[scale=.5]{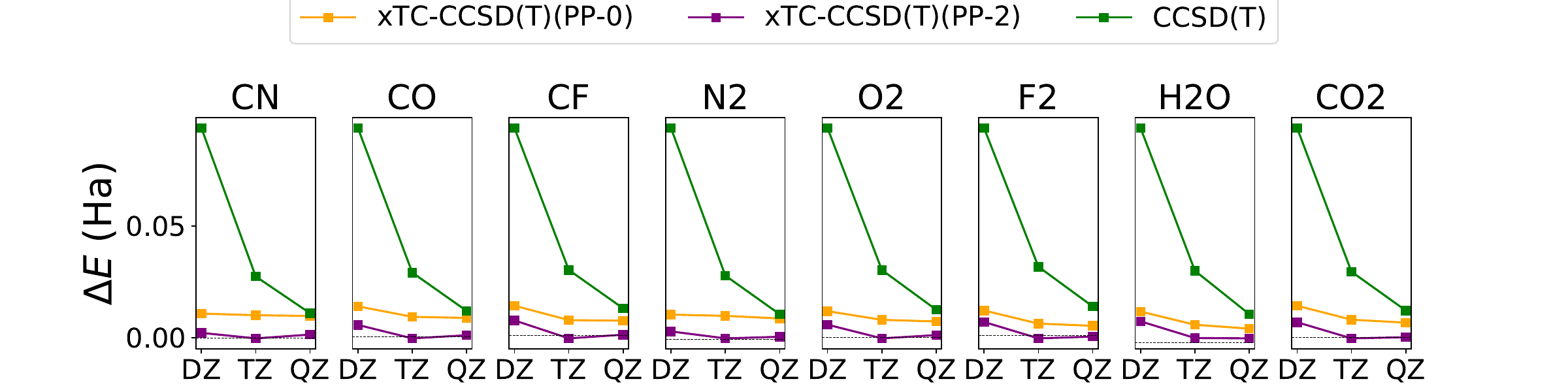}
    \end{subfigure}\vspace*{1cm}
    \begin{subfigure}{0.6\textwidth}
        \centering
        \caption{ccECPs:} 
        \hspace*{-5cm}
    \includegraphics[scale=.5]{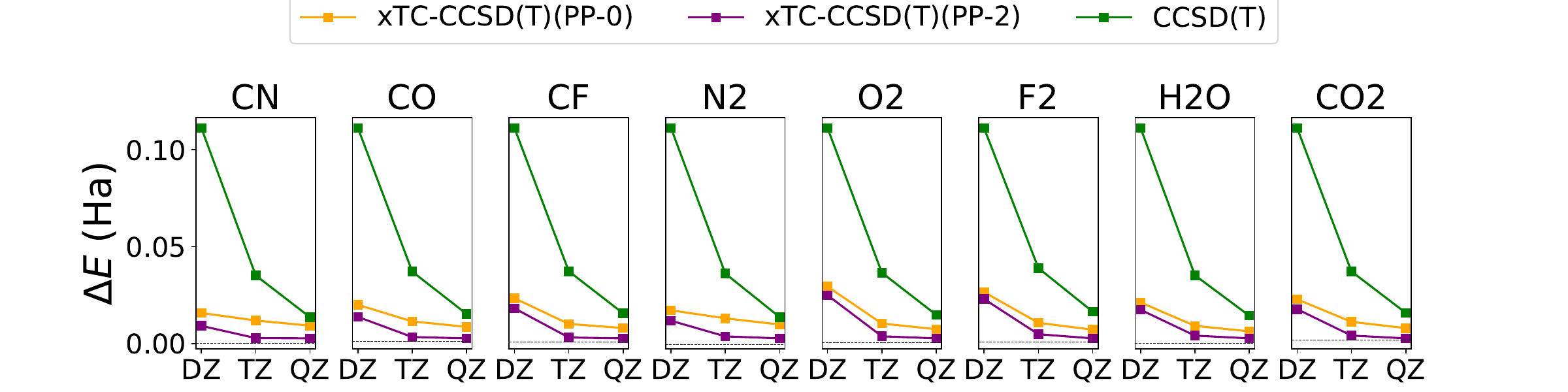}
    \end{subfigure}
    \caption{\label{fig: molecule_total_energies} Energies of the
     molecules CN, CO, CF, N$_2$, O$_2$, F$_2$, H$_2$O, and CO$_2$  with eCEPPs (a) and ccECPs (b) 
     as a function of the basis set. The energies are relative to  CCSD(T) 
     CBS energies as
      $\Delta E=E -  E_\text{CBS}^\text{TZ-QZ}(\text{CCSD(T)})$, see Eq. (\ref{eq:extrapolated energy1}). 
      CCSD(T) (green), xTC-CCSD(T)(PP-$0$) (yellow), and xTC-CCSD(T)(PP-$2$) (violet) energies are shown.}
\end{figure}

Figure \ref{fig: molecule_total_energies} shows the energies 
of the molecules CN, CO, CF, N\textsubscript{2}, O\textsubscript{2}, F\textsubscript{2}, H\textsubscript{2}O, and CO\textsubscript{2}, obtained with eCEPPs 
(a) and ccECPs (b) with CCSD(T) and xTC-CCSD(T)(PP-$n$) methods with $n=0$ and 
$n=2$. The energies are displayed
relative to $E_\text{CBS}^\text{TZ-QZ}(\text{CCSD(T)})$. The results are shown 
as a function of the basis set. 

The xTC energies are always below CCSD(T) energies at all basis sets.
Evaluation 
of the PP commutators decreases the energies. 
Unlike with some atoms and ions, xTC-CCSD(T)(PP-$0$) is still lower in energy
than CCSD(T) for the molecules. 
The increase of the basis set size has a very small effect to the xTC-CCSD(T) 
energies compared to the basis-set dependence of CCSD(T) energies.

\subsubsection{F12 atomization energies}

\begin{figure}[H]
    \includegraphics[scale=.4]{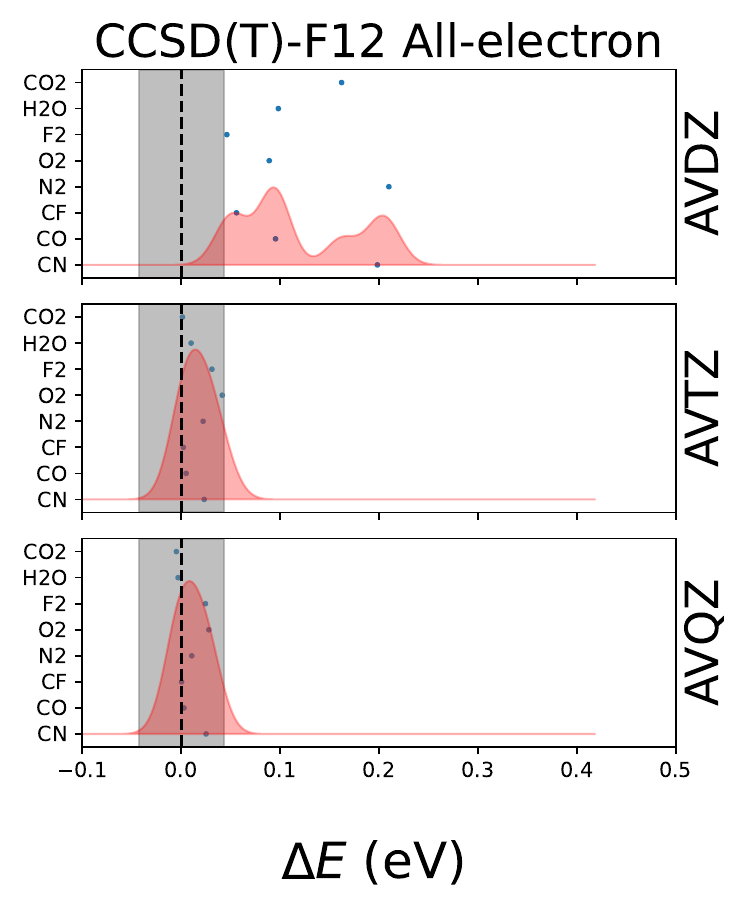}
    \includegraphics[scale=.4]{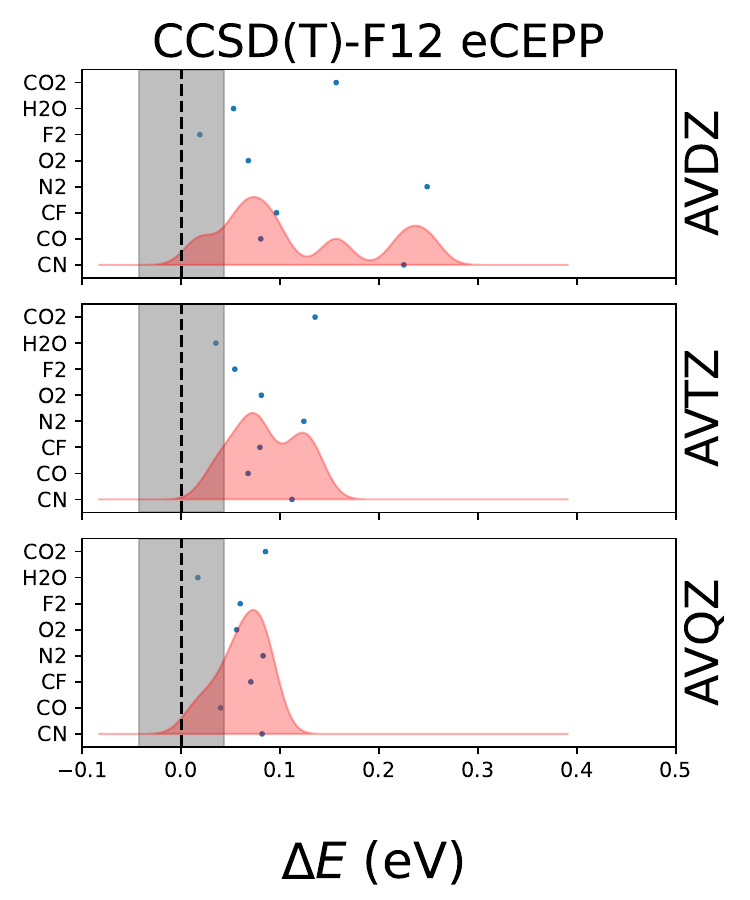}
    \includegraphics[scale=.4]{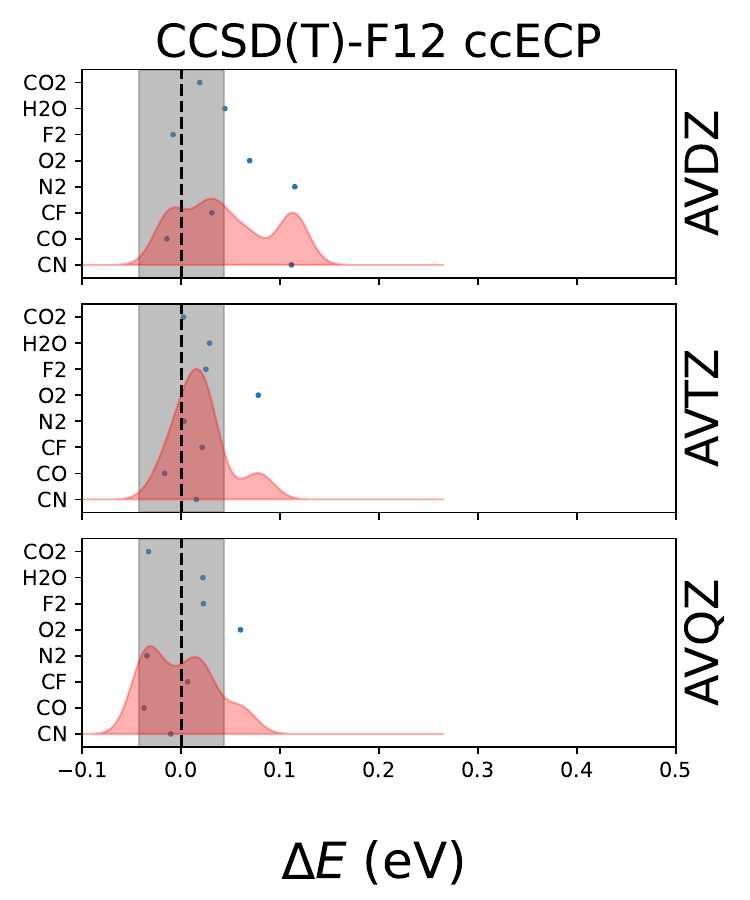}
    \caption{\label{fig: f12 AOs}Atomization energies of the
     molecules CN, CO, CF, N\textsubscript{2}, O\textsubscript{2}, F\textsubscript{2}, H\textsubscript{2}O, and CO\textsubscript{2}, evaluated as 
     $E_\text{at}=\sum E_\text{atom}-E_\text{mol}  $, calculated with CCSD(T)-F12
     method. The energies are presented relative to the results in the HEAT database,
     so that the presented 
     energies are $\Delta E = E_\text{HEAT} - E_\text{at}$. Results are obtained using 
     AVXZ basis sets, with X=D,T,Q (see labels on right). Results from all-electron (top),
     eCEPP (middle), and ccECP (bottom) calculations are shown. The grey shaded region 
     denotes chemical accuracy. The red shading represents a sum of gaussians centered 
     at the atomization energy discrepancies of each molecule, 
     with sigmas set so that equidistantly placed gaussians over the interval
     between maximum and minimum of any calculation with a given core treatment (AE, eCEPP, ccECP)
     would have nearest-neighbour distance of $4$ sigma. 
    }
\end{figure}

Figure \ref{fig: f12 AOs} shows the atomization energies of the molecules
CN, CO, CF, N\textsubscript{2}, O\textsubscript{2}, F\textsubscript{2}, H\textsubscript{2}O, and CO\textsubscript{2}, 
calculated with CCSD(T)-F12. The results are shown relative to the HEAT database
\cite{tajti2004}. The results are obtained with the AVXZ basis sets,
with X=D,T,Q. The results are shown for all-electron, eCEPP, and ccECP calculations.

The all-electron CCSD(T)-F12 method shows excellent accuracy.
With eCEPPs the accuracy is clearly worse, but then again with ccECPs all atoms 
are within chemical accuracy, except N\textsubscript{2}, which has a discrepancy of
$ \sim 50$ meV. 

Table \ref{tab:F12 atomisation_errors} shows the MAE and RMS of the atomization energies of the 
molecules CN, CO, CF, N\textsubscript{2}, O\textsubscript{2}, F\textsubscript{2}, H\textsubscript{2}O, and CO\textsubscript{2},
obtained with CCSD(T)-F12 and MRCI-F12 methods. The errors are averaged over the molecules studied,
and are shown separately for each basis set and core treatment. CCSD(T)-F12 is chemically accurate
in AVTZ and AVQZ basis sets with all-electron calculations and with ccECPs both in terms 
of MAE and RMS. 

The eCEPPs do not provide chemical accuracy. An interesting observation is that with
eCEPPs the MAE and RMS decrease with increasing basis set accuracy, but that the best MAE and RMS with
ECPs is obtained with the AVTZ basis set, and the AVQZ basis set is worse in MAE and almost equivalent
in RMS. This decrease 
in accuracy with ccECPs when moving from triple to quadruple zeta basis was already seen
 with the ionisation energies of the first row elements and with the xTC-CCSD(T) method. 

\begin{table}[!h]
    \centering
    \caption{\label{tab:F12 atomisation_errors} Mean average and root-mean square errors (MAE and RMS) of 
    the CCSD(T)-F12 atomization energies of the molecules studied. The all-electron values, as well as ccECP 
    and eCEPP pseudopotential results are shown. }

\begin{tabular}{lccccccccc}
    \toprule
    & \multicolumn{3}{c}{\textbf{All-electron}} & \multicolumn{3}{c}{\textbf{eCEPP}} & \multicolumn{3}{c}{\textbf{ccECP}} \\
    \cmidrule(lr){2-4} \cmidrule(lr){5-7} \cmidrule(lr){8-10}
    \textbf{Quantity} & \textbf{avdz} & \textbf{avtz} & \textbf{avqz} & \textbf{avdz} & \textbf{avtz} & \textbf{avqz} & \textbf{avdz} & \textbf{avtz} & \textbf{avqz} \\
    \midrule
    \textbf{MAE}   & 0.1133 & 0.0193 & 0.0135 & 0.1184 & 0.0861 & 0.0616 & 0.0515 & 0.0238 & 0.0283 \\
    \textbf{RMS}   & 0.1284 & 0.0235 & 0.0175 & 0.1417 & 0.0921 & 0.0656 & 0.0652 & 0.0326 & 0.0325 \\
    \bottomrule
\end{tabular}
\end{table}

\subsubsection{Atomization energies with transcorrelation}

\begin{figure}[H]
    \textbf{eCEPP Atomization energies with AVXZ}\newline
    % \begin{subfigure}{0.9\textwidth}
    %     \centering
    % %\includegraphics[scale=.6]{molecule_atomisation_energies.pdf}
    % \includegraphics[scale=.6]{atomisation_ecepp_evelin_dcsd.pdf}
    % \end{subfigure} \\
    \begin{subfigure}{0.9\textwidth}\hspace*{-1cm}
        \centering
    \includegraphics[scale=.7]{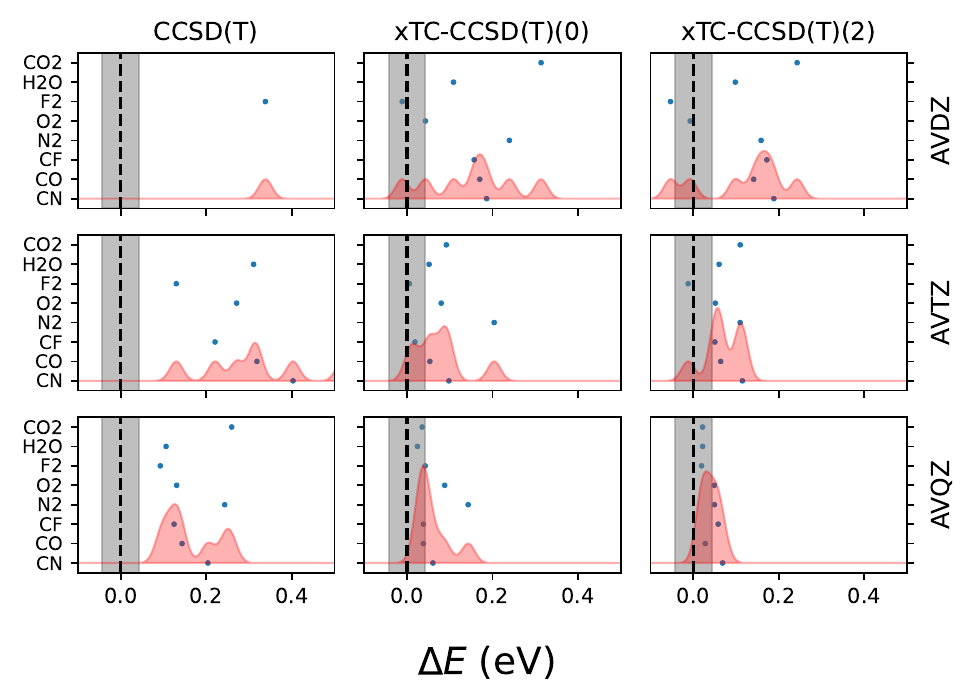}
    \end{subfigure}\newline
    \begin{subfigure}{0.9\textwidth}\hspace*{-1cm}
        \centering
    \includegraphics[scale=.7]{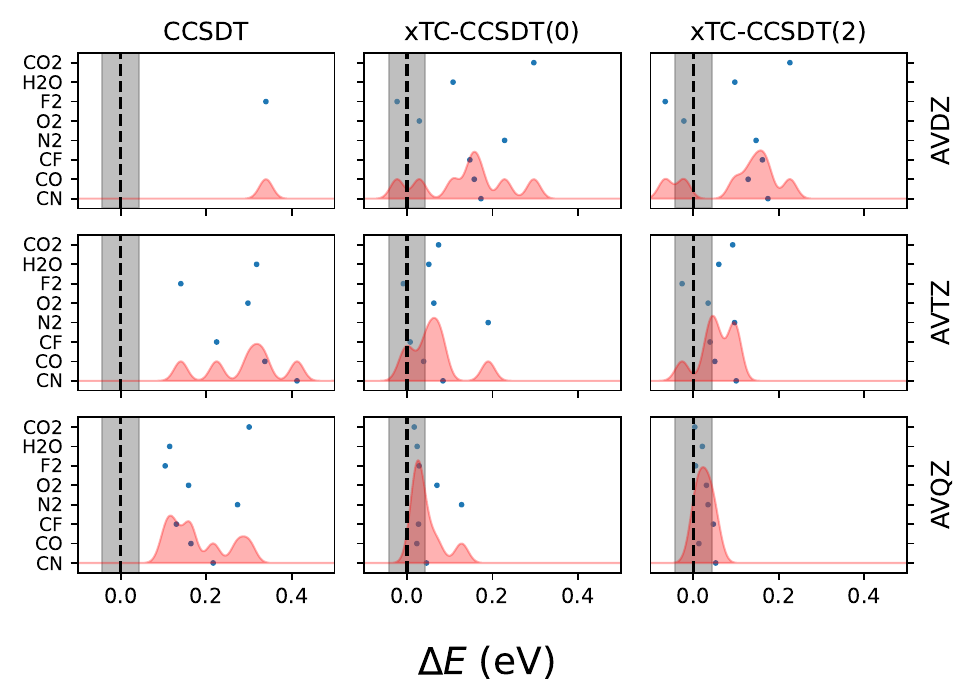}
    \end{subfigure} \\
    % \begin{subfigure}{0.9\textwidth}
    %     \centering
    % %\includegraphics[scale=.6]{molecule_atomisation_energies_lm.pdf}
    % \includegraphics[scale=.6]{atomisation_ecepp_evelin_dc_ccsdt.pdf}
    % \end{subfigure}
    \caption{\label{fig: ecepp_atomisation_energies}Atomization energies of a test set of 
    molecules, 
    evaluated with eCEPPs as 
     $ E_\text{at}=\sum E_\text{atom}-E_\text{mol}$. The energies are 
     presented relative to the results in the HEAT database, so that the presented 
     energies are $\Delta E = E_\text{HEAT} - E_\text{at}$. Calculations are shown without 
     TC (left column),and  with xTC both without and with 2 PP commutator
     evaluations (\{ \textit{method}\}(0) and \{ \textit{method}\}(2), respectively, method 
     being either CCSD(T) or CCSDT).
     The grey shaded region 
     denotes chemical accuracy. The red shading represents a sum of gaussians centered 
     at the atomization energy discrepancies of each molecule, 
     with sigmas set so that equidistantly placed gaussians over the interval
     between maximum and minimum of a given level of theory
     would have nearest-neighbour distance of $4$ sigma. To improve presentation
     and to better compare with CCSD(T)-F12 results, we are only showing the region
     $-0.1 < \Delta E < 0.5$eV, which leaves some of the energies of non-transcorrelated methods outside 
     of the range in AVDZ and AVTZ basis sets.}
    
\end{figure}

\begin{figure}[H]
    \textbf{ccECP Atomization energies with AVXZ}\newline
    % \begin{subfigure}{0.9\textwidth}
    %     \centering
    % %\includegraphics[scale=.6]{molecule_atomisation_energies.pdf}
    % \includegraphics[scale=.6]{atomisation_ecp_evelin_dcsd_aug.pdf}
    % \end{subfigure} \\
    \begin{subfigure}{0.9\textwidth}\hspace*{-2cm}
        \centering
   \includegraphics[scale=.7]{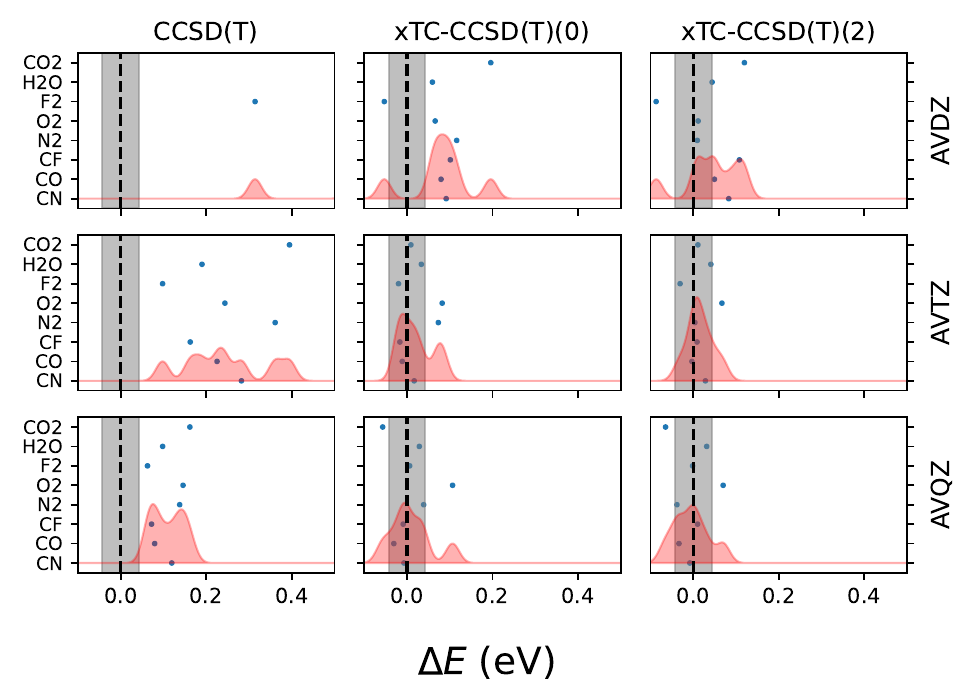}
    \end{subfigure}\newline
    \begin{subfigure}{0.9\textwidth}\hspace*{-2cm}
        \centering
    \includegraphics[scale=.7]{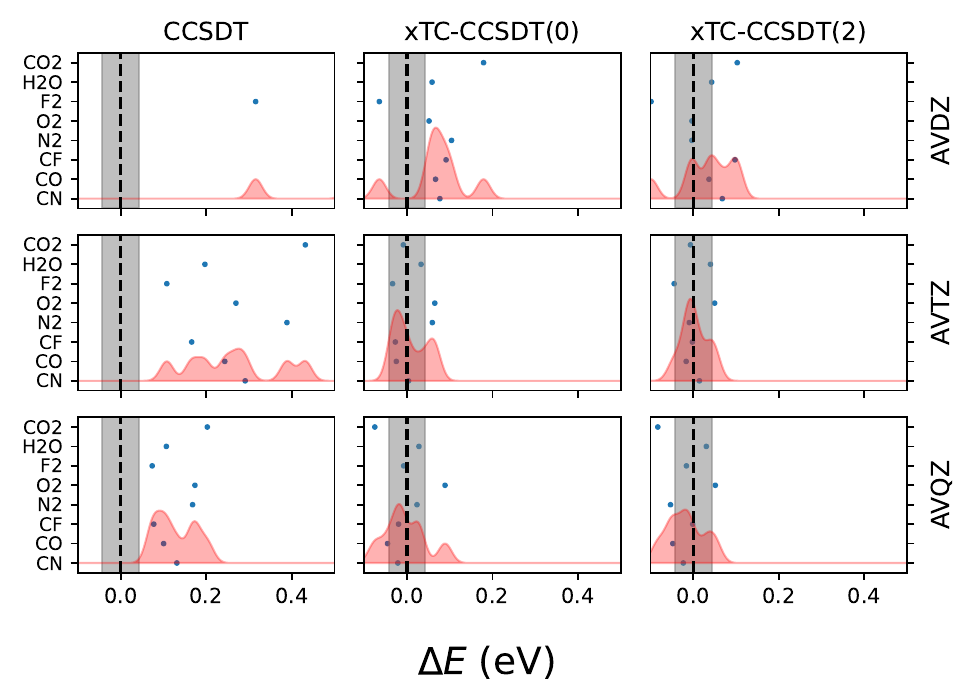}
    \end{subfigure} \\
    % \begin{subfigure}{0.9\textwidth}
    %     \centering
    % %\includegraphics[scale=.6]{molecule_atomisation_energies_lm.pdf}
    % \includegraphics[scale=.6]{atomisation_ecp_evelin_dc_ccsdt_aug.pdf}
    % \end{subfigure}
    \caption{\label{fig: ccECP_atomisation_energies with augmentation} Results evaluated as in 
    Fig. \ref{fig: ecepp_atomisation_energies}, but with ccECPs and 
    AVXZ (X=D,T,Q) basis sets. }
\end{figure}

We calculated the atomization energies of the molecules CN, CO, CF, N\textsubscript{2}, 
O\textsubscript{2}, F\textsubscript{2}, H\textsubscript{2}O, and CO\textsubscript{2}, 
using eCEPPs and ccECPs with AVXZ basis sets (Figures~\ref{fig: ecepp_atomisation_energies} 
and \ref{fig: ccECP_atomisation_energies with augmentation}). The results are 
presented as discrepancies relative to the HEAT values \cite{tajti2004}, employing  CCSD(T)
 and CCSDT \cite{elemcojl}. Both methods were
applied in three variants: non-TC, xTC without PP commutator evaluations, and xTC with 
two PP commutator evaluations.


The eCEPP results (Figure~\ref{fig: ecepp_atomisation_energies}) show consistent improvement in 
accuracy with basis set size. Both xTC variants outperform the 
non-TC calculations with both CCSD(T) and CCSDT, with PP commutator evaluations further enhancing 
accuracy. xTC-CCSDT(2) reaches chemical accuracy for all molecules studied.
The results with ccECPs show best performance with xTC and ECP 
commutator evaluations in the AVTZ basis, where the results 
are even more accurate than with eCEPPs in AVQZ basis. However,
the ccECP results in AVQZ basis, despite being within chemical accuracy
(with CO2 slightly outside of the chemical accuracy regime), are
slightly worsened from AVTZ results.

Table~\ref{table: MAE and RMS of Atomization energies evaluated with different methods, basis sets, and PPs.} 
summarizes the mean absolute errors (MAE) and root mean square errors (RMS) for all combinations of PP type, 
method, basis set, and theory. In terms of MAE and RMS, non-transcorrelated methods remain far from chemical accuracy even 
with the AVQZ basis set. In contrast, transcorrelated methods with eCEPPs, AVQZ basis sets, and two PP commutator 
evaluations achieve chemical accuracy across all studied methods. With ccECPs, xTC methods with two 
PP commutator evaluations are within or very close to chemical accuracy with AVQZ basis, and 
within chemical accuracy using the AVTZ basis set. The fact that we
see best results with AVTZ basis, and not with AVQZ basis, when using ccECPs both with F12 and
transcorrelated methods hints that the non-monotonic increase in accuracy with basis set resolution
is a feature of the ccECPs.

\begin{table}[h!]
    \centering
    \caption{\label{table: MAE and RMS of Atomization energies evaluated with different methods, basis sets, and PPs.} 
    Mean Absolute Errors (MAE) and Root 
    Mean Square Errors (RMS) for Different Methods.}
    
    \textbf{MAE}
    
    \begin{tabular}{lcccccccccc}
        \toprule
        \textbf{Method} & \multicolumn{3}{c}{\textbf{eCEPP}} & \multicolumn{3}{c}{\textbf{ccECP}} \\
        \cmidrule(lr){2-4} \cmidrule(lr){5-7}
                         & \textbf{avdz} & \textbf{avtz} & \textbf{avqz} & \textbf{avdz} & \textbf{avtz} & \textbf{avqz} \\
        \midrule
        % \textbf{DCSD}         & 0.923 & 0.441 & 0.275 & 1.082 & 0.468 & 0.262 & 0.801 & 0.346 & 0.221 \\
        % \textbf{xTC-DCSD (PP-$0$)}  & 0.149 & 0.056 & 0.039 & 0.273 & 0.056 & 0.041  & 0.090 & 0.026 & 0.036\\
        % \textbf{xTC-DCSD (PP-$2$)}  & 0.140 & 0.057 & 0.026 & 0.261 & 0.048 & 0.050 & 0.078 & 0.035 & 0.048 \\
        % \midrule
        \textbf{CCSD(T)}       & 0.900 & 0.340 & 0.163 &  0.773 & 0.245 & 0.101 \\
        \textbf{xTC-CCSD(T) (PP-$0$)} & 0.154 & 0.075 & 0.059 & 0.095 & 0.033 & 0.036 \\
        \textbf{xTC-CCSD(T) (PP-$2$)} & 0.133 & 0.072 & 0.039 & 0.064 & 0.024 & 0.032 \\
        \midrule
        \textbf{CCSDT}         & 0.904 & 0.358 & 0.183 & 0.776 & 0.262 & 0.129  \\
        \textbf{xTC-CCSDT (PP-$0$)}  & 0.145 & 0.065 & 0.045 & 0.087 & 0.032 & 0.039 \\
        \textbf{xTC-CCSDT (PP-$2$)}  & 0.128 & 0.062 & 0.026 & 0.057 & 0.023 & 0.038 \\
        % \midrule
        % \textbf{DC-CCSDT}      & 0.889 & 0.339 & 0.164 & 1.043 & 0.366 & 0.151  & 0.760 & 0.243 & 0.113 \\
        % \textbf{xTC-DC-CCSDT (PP-$0$)} & 0.143 & 0.063 & 0.043 & 0.272 & 0.064 & 0.045 & 0.085 & 0.032 & 0.040\\
        % \textbf{xTC-DC-CCSDT (PP-$2$)} & 0.126 & 0.060 & 0.024 & 0.266 & 0.052 & 0.045 & 0.056 & 0.024 & 0.040 \\
        \bottomrule
    \end{tabular}

    \vspace{0.5cm}
    \textbf{RMS}
    
    \begin{tabular}{lcccccccccc}
        \toprule
        \textbf{Method} & \multicolumn{3}{c}{\textbf{eCEPP}} & \multicolumn{3}{c}{\textbf{ccECP}} \\
        \cmidrule(lr){2-4} \cmidrule(lr){5-7}
                         & \textbf{avdz} & \textbf{avtz} & \textbf{avqz} & \textbf{avdz} & \textbf{avtz} & \textbf{avqz} \\
        % \midrule
        % \textbf{DCSD}         & 1.003 & 0.477 & 0.297 & 1.126 & 0.493 & 0.276 & 0.872 & 0.374 & 0.237 \\
        % \textbf{xTC-DCSD (PP-$0$)}  & 0.180 & 0.069 & 0.044 & 0.355 & 0.063 & 0.049 & 0.109 & 0.031 & 0.042 \\
        % \textbf{xTC-DCSD (PP-$2$)}  & 0.156 & 0.067 & 0.037 & 0.345 & 0.055 & 0.057 & 0.090 & 0.040 & 0.057 \\
        \midrule
        \textbf{CCSD(T)}       & 0.969 & 0.366 & 0.174 & 0.833 & 0.262 & 0.115 \\
        \textbf{xTC-CCSD(T) (PP-$0$)} & 0.179 & 0.095 & 0.069 & 0.105 & 0.043 & 0.048 \\
        \textbf{xTC-CCSD(T) (PP-$2$)} & 0.151 & 0.079 & 0.043 & 0.075 & 0.032 & 0.040 \\
        \midrule
        \textbf{CCSDT}         & 0.974 & 0.385 & 0.195 & 0.837 & 0.281 & 0.137  \\
        \textbf{xTC-CCSDT (PP-$0$)}  & 0.169 & 0.084 & 0.057 & 0.095 & 0.038 & 0.047 \\
        \textbf{xTC-CCSDT (PP-$2$)}  & 0.141 & 0.068 & 0.031 & 0.069 & 0.030 & 0.045\\
        % \midrule
        % \textbf{DC-CCSDT}      & 0.958 & 0.365 & 0.175 & 1.081 & 0.383 & 0.158 & 0.820 & 0.261 & 0.121 \\
        % \textbf{xTC-DC-CCSDT (PP-$0$)} & 0.167 & 0.083 & 0.055 & 0.358 & 0.081 & 0.055 & 0.093 & 0.038 & 0.048 \\
        % \textbf{xTC-DC-CCSDT (PP-$2$)} & 0.139 & 0.066 & 0.029 & 0.349 & 0.064 & 0.055 & 0.067 & 0.029 & 0.047 \\
       \bottomrule
    \end{tabular}
\end{table}

\subsection{Dissociation energies}

\subsubsection{N\textsubscript{2} \label{subsubsection: N2}}

% \begin{figure}[!h]
%     \centering
%     \hspace*{-2cm}
%     %\includegraphics[scale=.6]{N2_dissociation_mrci.pdf}
%     \includegraphics[scale=.6]{N2_dissociation_mrci.pdf}
%     \caption{N2 FCIQMC energies as a function of bond lenght with eCEPPS, computed in AVDZ (top row), AVTZ (middle row)
%     and AVQZ (bottom row) basis sets. Results are presented as differences to AVQZ MRCI-F12 energies. Results without TC
%     (brown), and with TC with $0$ (green), $1$ (yellow), and $2$ (violet) ccECP commutator evaluations are shown. The grey shaded region 
%     denotes chemical accuracy to MRCI-F12 results. The dotted vertical line points to the equilibrium bond length of N2.
%     }
%     \label{fig: N2 dissociation mrci}
% \end{figure}

\begin{figure}[H]
    \centering
    \hspace*{-2cm}
    \includegraphics[scale=.6]{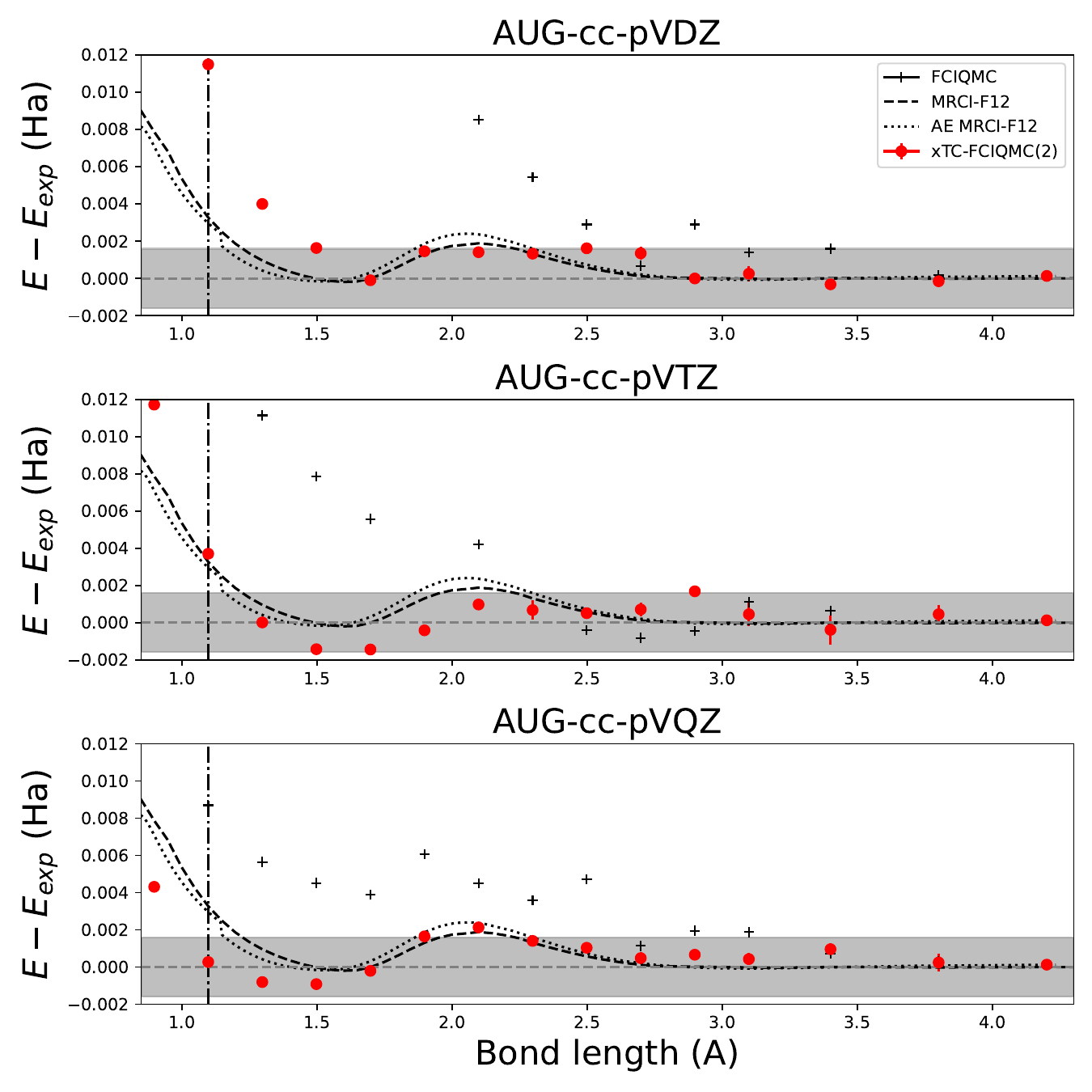}
    \caption{N\textsubscript{2} FCIQMC (black crosses) and xTC-FCIQMC(PP-$2$) (red circles) dissociation curves 
    as a function of bond length with eCEPPS, 
    computed in AVDZ (top row), AVTZ (middle row) and AVQZ 
    (bottom row) basis sets. MRCI-F12 energies are evaluated in AVQZ basis set
    in all of the images, and both eCEPP (dashed black) and all-electron MRCI-F12 
    (dotted black) curves are shown. MRCI-F12 results are evaluated in the
    AVQZ basis set in all of the images. 
    Results are presented as differences to 
    experimental dissociation curve from Ref. \cite{leroy2006}, taking the overlap 
    between theory and experiment to be at $r=4.2$ \AA \ .The grey shaded region 
    denotes chemical accuracy with respect to experiment. The dash-dotted vertical line points 
    to the equilibrium bond length of N\textsubscript{2} of $r=1.098$ \AA.
    }
    \label{fig: N2 dissociation exp}
\end{figure}

Figure \ref{fig: N2 dissociation exp} shows energy differences between experimental and theoretical
evaluations of the N\textsubscript{2} dissociation curve. Energy differences are presented as a function of bond length.
The experimental curve is taken from Ref. \cite{leroy2006}.
The theoretical curves are evaluated with FCIQMC and MRCI methods.
We used eCEPPS and did the calculations in AVDZ,
AVTZ, and AVQZ basis sets using FCIQMC. We show results
for FCIQMC and xTC-FCIQMC(PP-$2$). MRCI-F12 results are only evaluated in the
AVQZ basis set. The theoretical results are shifted to overlap with 
experiment at $r=4.2$ \AA, a length which corresponds to the system of two isolated 
nitrogen atoms. 

\clearpage

Because of the strong correlations and change of spin state involved in dissociation of the N\textsubscript{2}, 
the Hartree-Fock wave function is a poor reference for the Jastrow factor optimization. 
Hence we used a method developed recently to tackle this problem \cite{haupt}.
First, we took the $100$ most populated determinants from a non-transcorrelated FCIQMC calculation,
and used them for constructing a trial wave function for VMC, and optimized the
Jastrow factor with the VMC method before using it for preparing the transcorrelated
Hamiltonian. In the subsequent xTC phase we used the reduced density matrix of the 
multideterminant wave function used in the optimisation.

Figure \ref{fig: N2 dissociation exp} shows that the regular (non-transcorrelated) FCIQMC does not 
achieve chemical accuracy with 
respect to basis-set-error at near-equilibrium bond lengths. xTC-FCIQMC(PP-$2$)
underestimates the equilibrium energy 
at triple-zeta basis by about 4mH, but has an excellent match in quadruple zeta-basis, a feature already seen with 
xTC coupled cluster methods and eCEPPs in Fig. \ref{fig: ecepp_atomisation_energies}. The xTC-FCIQMC(PP-$2$)
method in quadruple zeta-basis 
obtains chemical accuracy everywhere except at the most compressed bond length (below 1\AA),  and at $r=2.098$ \AA \ , 
where, however, the stochastic error of the result overlaps with chemical accuracy regime. 

We tested the size-consistency by calculating the difference between the FCIQMC energies of two isolated nitrogen atoms and the energy of the N\textsubscript{2} molecule at the longest bond distance studied, $\Delta E = 2E_{\text{N}} - E_{\text{N}_2}(r=4.2\AA)$. This test was done in AVDZ basis, and resulted in $\Delta E=-1.5(2)$mHa. This small size-inconsistency can be traced to the use of the combined Jastrow treatment (see discussion after Eq. (\ref{eq: j-factor})) together with the xTC approximation, since the latter approximates the effect of the three-body interactions introduced by the commutators of the kinetic energy operator and the Jastrow factor. In the combined Jastrow approach the one-body terms are folded into 2-body Jastrow terms, which in turn leads to additional 3-body interactions (approximated in the xTC treatment). This combination leads to size-inconsistency. This size-consistency error can be entirely eliminated by using the separated Jastrow treatment (See Supplementary material for the separate treatment of PP commutators). We verified explicitly with further calculations that the xTC approximation, in the separate Jastrow treatment, does not incur a size-consistency error. Because we checked that in equilibrium geometries the separate and combined treatments yielded the same results, the overall error in this study due to the aforementioned size-inconsistency is on the order of 1-2mHa. The best workflow for future studies will require further investigation of Jastrow cutoffs and the use of the separate Jastrow treatment.

The MRCI-F12 results are in good agreement with the experimental curve at longer distances, 
but overestimate the energy at equilibrium distance. The correspondence between the
eCEPP and all-electron curves indicate that the eCEPPs introduce almost no error into the simulation. 

\subsubsection{F\textsubscript{2}}

\begin{figure}[H]
    \centering
    \hspace*{-2cm}
    \includegraphics[scale=.6]{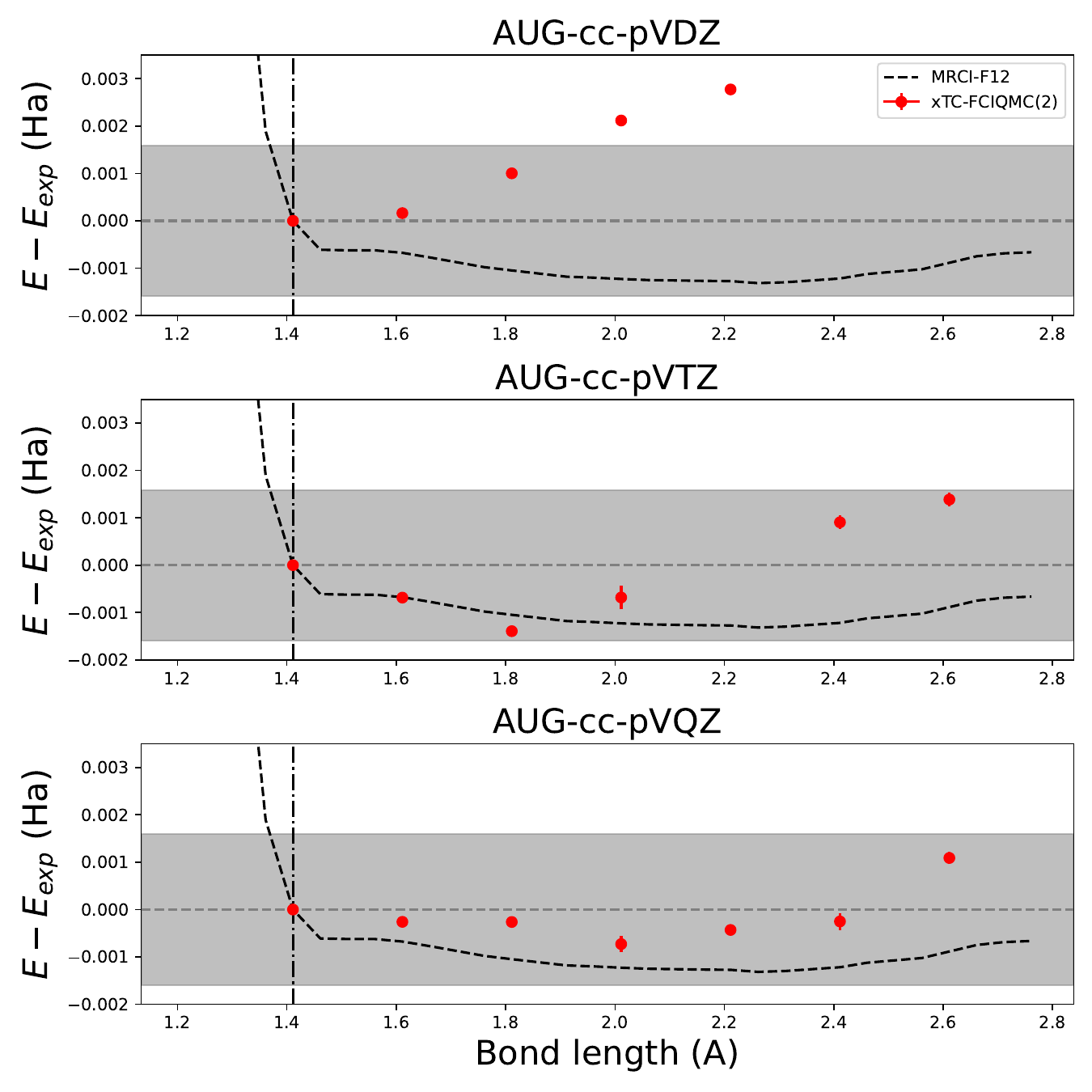}
    \caption{F\textsubscript{2} xTC-FCIQMC(PP-$2$) and MRCI-F12 energies as a function of bond length with eCEPPS, 
    computed in AVDZ (top row), AVTZ (middle row) and AVQZ (bottom 
    row) basis sets. MRCI-F12 energies are the same in all plots, and are evaluated in 
    AVQZ basis set. Results are presented as differences to experimental 
    dissociation curve from Ref. \cite{colbourn1976}. The grey shaded region denotes the chemical 
    accuracy with respect to experiment. The dotted vertical line points to the equilibrium bond 
    length of F\textsubscript{2} of $1.4112$ \AA.
    }
    \label{fig: F2 dissociation exp}
\end{figure}

Figure \ref{fig: F2 dissociation exp} shows the dissociation error
curve of F\textsubscript{2}, showing the difference between MRCI-F12 and xTC-FCIQMC(PP-$2$)  theoretical methods
and experiment.
xTC-FCIQMC(PP-$2$) results are evaluated with eCEPPS and computed in AVDZ, AVTZ, and 
AVQZ basis sets. The MRCI-F12 results are evaluated in the AVQZ basis set.
The experimental curve is taken from Ref. \cite{colbourn1976}.
Because the experimental data extends only to $r\sim 2.8$ \AA, the theoretical results are
shifted to overlap with experiment at the equilibrium bond length of $r=1.4118$ \AA.
 
The Jastrow factors for
F\textsubscript{2} are optimized with only the RHF determinant in the VMC method. This 
proves to be already highly accurate, as for F\textsubscript{2} only a single bond is broken 
and less correlation is involved in the dissociation. 

The xTC-FCIQMC(PP-$2$) results at triple zeta and higher basis sets are very accurate, 
with a sub-mHa error along the whole range of the plot with AVQZ basis set.  MRCI-F12
are also chemically accurate with the AVQZ basis set, but the error is larger.

\section{Conclusions}

We have presented a study of transcorrelated theory under PP approximations. 
The study included the derivation of the PP commutators needed for evaluating
the transcorrelated Hamiltonian. The algorithms were implemented using an in-house code
TCHINT, that
was subsequently used to evaluate the accuracy of the transcorrelated methods with
PPs.

The accuracy of the method was evaluated by estimating the ionisation potentials of atoms 
in the first row, the atomization energies of a test set of molecules, and the dissociation
curves of N\textsubscript{2} and F\textsubscript{2}. The results were compared to the HEAT database, experimental
data, and MRCI-F12 results.

xTC-CCSD(T)(PP-$2$) provided chemical accuracy for the ionisation energies of the first row atoms
with both eCEPPs and ccECPs. For the atomization energies, all of the coupled cluster levels of theory
with $2$ PP commutator evaluations
reached chemical accuracy with eCEPPs and AVQZ basis set. With ccECPs, chemical accuracy 
was reached with AVTZ basis set, while with AVQZ basis the RMS of the methods was slightly 
outside the chemical accuracy regime. 

Atomization energy calculations
with CCSD(T)-F12 showed the same trend of reduced accuracy when moving from AVTZ to AVQZ basis. 
This leads to the conclusion that with explicitly correlated methods, the ccECPs have some problems 
with augmented basis set convergence, but that the results are very accurate already at 
triple-zeta level.  The F12 methods were found 
to work better with ccECPs than eCEPPs. 

Finally, the dissociation curves of N\textsubscript{2} and F\textsubscript{2} were evaluated with xTC-FCIQMC
and MRCI-F12 methods. The xTC-FCIQMC results were very accurate with the AVQZ basis set, and apart 
from the compressed distance region of N\textsubscript{2}, chemical accuracy was reached. 

The results of this study show that the transcorrelated methods are very accurate with PPs,
and that the accuracy is comparable to the all-electron results. The evaluation of the PP commutators
is essential for the accuracy of the method. It is possible that optimisation of the Jastrow 
factor without the presence of the core electrons allows a more targeted TC simulation, focusing 
on the valence electrons, a feature than can provide useful in the future development of the
method.

The theory of transcorrelation with PPs can help bringing the applicability of the TC 
method to a wider range of systems. Calculations with larger system sizes can benefit 
from the reduced variance, making the Jastrow optimizations more feasible. Applications with 
heavier atoms, such as transition metals, would be an interesting future direction. 
Crucially, the methods presented in this work can help in the development of TC theory 
towards periodic solid-state systems, possibly even in the plane-wave basis. Directions 
towards application of TC theory with PPs in embedding models for systems such as solid-state
defects are also currently under investigation.

\section*{Supporting Information}
Supporting Information.  
• Optimized Jastrow factor parameters for atoms and molecules studied in this work.  
• Derivations of pseudopotential commutator formulas (separate Jastrow treatment).  
• FCIDUMP files containing transcorrelated second-quantized Hamiltonians for Be in 
  AVDZ basis set,  with and without the PP commutator evaluations.

\clearpage

%\bibliography{mybib} % Ensure your BibTeX file is named 'mybib.bib'
%\bibliographystyle{achemso}
\providecommand{\latin}[1]{#1}
\makeatletter
\providecommand{\doi}
  {\begingroup\let\do\@makeother\dospecials
  \catcode`\{=1 \catcode`\}=2 \doi@aux}
\providecommand{\doi@aux}[1]{\endgroup\texttt{#1}}
\makeatother
\providecommand*\mcitethebibliography{\thebibliography}
\csname @ifundefined\endcsname{endmcitethebibliography}
  {\let\endmcitethebibliography\endthebibliography}{}

\end{document}